\RequirePackage{rotating}
\documentclass[useAMS, usenatbib]{mn2e}
\usepackage{color}
\usepackage{epsfig}
\usepackage{epstopdf}
\usepackage{amsmath, amssymb}
\usepackage{hyperref}
\usepackage{wasysym} 
\usepackage{subfigure}
\usepackage{multirow}
\usepackage{lscape}
\usepackage{lipsum}
\usepackage{ulem}
\usepackage{mn2e-breakabs}
\usepackage{setspace}  

\newlength{\plotwidth}
\newlength{\fullwidth}
\newcommand{\hMpc}{h^{-1}{\rm\;Mpc}}

\def\vc#1{{\bf#1}}

\pagerange{\pageref{firstpage}--\pageref{lastpage}} \pubyear{2013}

\def\LaTeX{L\kern-.36em\raise.3ex\hbox{a}\kern-.15em
    T\kern-.1667em\lower.7ex\hbox{E}\kern-.125emX}

\title[Constraining the relative velocity effect using BOSS DR12]{Constraining the relative velocity effect using the Baryon Oscillation Spectroscopic Survey}
\author[Beutler et al.]
{\parbox{\textwidth}{Florian Beutler$^{1}$\thanks{E-mail: \texttt{florian.beutler@port.ac.uk}}, Uro\v s Seljak$^{2,3}$, Zvonimir Vlah$^{4,5}$}\vspace{0.4cm}\\
\parbox{\textwidth}{
$^{1}$Institute of Cosmology \& Gravitation, Dennis Sciama Building, University of Portsmouth, Portsmouth, PO1 3FX, UK\\
$^{2}$Lawrence Berkeley National Lab, 1 Cyclotron Rd, Berkeley CA 94720, USA\\
$^{3}$Department of Physics, University of California Berkeley, CA 94720, USA\\
$^{4}$Stanford Institute for Theoretical Physics and Department of Physics, Stanford University, Stanford, CA 94306, USA\\
$^{5}$Kavli Institute for Particle Astrophysics and Cosmology, SLAC and Stanford University, Menlo Park, CA 94025, USA\vspace{-0.5cm}
}}

\begin{document}
\maketitle

\begin{abstract}
We analyse the power spectrum of the Baryon Oscillation Spectroscopic Survey (BOSS), Data Release 12 (DR12) to constrain the relative velocity effect, which represents a potential systematic for measurements of the Baryon Acoustic Oscillation (BAO) scale. The relative velocity effect is sourced by the different evolution of baryon and cold dark matter perturbations before decoupling. Our power spectrum model includes all $1$-loop redshift-space terms corresponding to $v_{\rm bc}$ parameterised by the bias parameter $b_{v^2}$. We also include the linear terms proportional to the relative density, $\delta_{\rm bc}$, and relative velocity dispersion, $\theta_{\rm bc}$, which we parameterise with the bias parameters $b^{\rm bc}_{\delta}$ and $b^{\rm bc}_{\theta}$.
Our data does not support a detection of the relative velocity effect in any of these parameters. Combining the low and high redshift bins of BOSS, we find limits of $b_{v^2} = 0.012 \pm 0.015\;(\pm 0.031)$, $b^{\rm bc}_{\delta} = -1.0 \pm 2.5\;(\pm 6.2)$ and $b^{\rm bc}_{\theta} = -114 \pm 55\;(\pm 175)$ with $68\%$ ($95\%$) confidence levels. These constraints  restrict the potential systematic shift in $D_A(z)$, $H(z)$ and $f\sigma_8$, due to the relative velocity, to $1\%$, $0.8\%$ and $2\%$, respectively. Given the current uncertainties on the BAO measurements of BOSS these shifts correspond to $0.53\sigma$, $0.5\sigma$ and $0.22\sigma$ for $D_A(z)$, $H(z)$ and $f\sigma_8$, respectively.
\end{abstract}


\section{introduction}
\label{sec:intro}

Measurements of the baryon acoustic scale in the distribution of galaxies have established themselves as one of the most powerful tools for precision cosmology~\citep{Eisenstein1998:astro-ph/9805239v1, Percival2001:astro-ph/0105252v1, Blake2003:astro-ph/0301632v2, Hu2003:astro-ph/0306053v1, Seo2003:astro-ph/0307460v1, Linder2003:astro-ph/0304001v1, Eisenstein2005:astro-ph/0501171, Cole2005:astro-ph/0501174v2, Beutler2011:1106.3366v1, Blake2011:1108.2635v1, Alam2016:1607.03155v1}. With the most recent measurements of the BAO scale in the BOSS survey we have now reached $1\%$ precision in two redshift bins~\citep{Alam2016:1607.03155v1, Beutler2016:1607.03149v1, Ross2016:1607.03145v1}. 

Given the fact that the BAO signal is located on very large scales, the impact of any late time non-linear evolution is small for these measurements and fairly simple perturbation theory based models can be used to extract the BAO scale~\citep{Crocce2007:0704.2783v2, Padmanabhan2008:0812.2905v3}. In the light of the next generation of galaxy redshift surveys like DESI~\citep{Schlegel2009:0904.0468v3} and Euclid~\citep{Laureijs2011:1110.3193v1}, which will reduce the uncertainties on these measurements by another order of magnitude, even small effects to the BAO scale can bias our cosmological constraints.

In this paper we investigate the relative velocity effect and its impact on anisotropic BAO and RSD measurements. The relative velocity effect is sourced by the photon pressure, which prevents baryon perturbations from growing before decoupling. This introduces a relative density $\delta_{\rm bc}$ and velocity divergence $\theta_{\rm bc}$ as well as a relative velocity $v_{\rm bc}$ between cold dark matter and baryonic matter. This relative velocity can shift the BAO scale and hence represents a possible systematic for future BAO measurements~\citep{Dalal2010:1009.4704v1, 1308.1401v2}. The relative velocity effect can impact the BAO scale, because it is sourced by the same physical effects, which imprinted the BAO scale itself, and hence this effect acts on the same scale. 

The relative velocity $v_{\rm bc}$ is about 30km/s at redshift 1000 and decays with $1/a$, reducing it to 0.03km/s at redshift zero. Therefore this effect is negligible at low redshift compared to the far larger virial velocities in galaxy groups and clusters. However, the relative velocity can prevent the condensation of baryons within the gravitational potential of the cold dark matter haloes and therefore impact early galaxy formation~\citep{Tseliakhovich2010:1005.2416v2, Dalal2010:1009.4704v1, Tseliakhovich2010:1012.2574v1, Fialkov2012:1212.0513v1, Naoz2012:1207.5515v3}. \citet{1308.1401v2} argue that the modulation of early, low-mass halos by the relative velocity, will effect the subsequent formation of high mass haloes observed today. Since these processes are not known in detail, the amplitude of the relative velocity effect cannot be predicted and must be constrained by the data.

In this paper we use the latest BOSS DR12 data to constrain the relative velocity effect. While such studies have been done before, there are several novel aspects to our analysis: (1) for the first time we include the advection term~\citep{1510.03554v2}, (2) beside $b_{v^2}$ we also set constraints on biasing by the density, $\delta_{\rm bc}$ and velocity divergence, $\theta_{\rm bc}$ (\citealt{Barkana2010:1009.1393v1,Schmidt2016:1602.09059v2}), (3) we include all relative velocity contributions up to 1-loop order including the redshift-space terms and (4) we quantify the potential shifts due to all three relative velocity contributions for the anisotropic BAO and RSD parameters.

This paper is organised as follows. We start with the introduction of the BOSS DR12 dataset in section~\ref{sec:data}. In section~\ref{sec:measurement} we present the power spectrum measurements, which we use for our analysis. In section~\ref{sec:model} we discuss the power spectrum model, which is based on perturbation theory and includes the relative velocity terms. In section~\ref{sec:systests} we introduce the mock catalogues which we use to test our model. In section~\ref{sec:analysis} we fit the BOSS measurements and constrain the relative velocity parameters. In section~\ref{sec:sysbias} we quantify the potential systematic uncertainty on the BAO scale given our constraints on the relative velocity parameters. We further discuss our results in section~\ref{sec:discussion} before concluding in section~\ref{sec:conclusion}.

The fiducial cosmological parameters, which are used to convert the observed angles and redshifts into co-moving coordinates and to generate linear power spectrum models as input for the power spectrum templates, follow a flat $\Lambda$CDM model with $\Omega_m=0.31$, $\Omega_bh^2=0.022$, $h=0.676$, $\sigma_8=0.824$, $n_s=0.96$, $\sum m_{\nu} = 0.06\,$eV and $r_s^{\rm fid} = 147.78\,$Mpc. These parameters are the fiducial cosmological parameters used for the BOSS DR12 data analysis and are close to the Planck 2015 cosmological constraints within $\Lambda$CDM.

\section{The BOSS DR12 dataset}
\label{sec:data}

BOSS, as part of SDSS-III~\citep{1101.1529v2,1208.0022v3} measured spectroscopic redshifts of $1\,198\,006$ galaxies making use of the SDSS multi-fibre spectrographs~\citep{1207.7326v2,1208.2233v2}. The galaxies are selected from multi-colour SDSS imaging~\citep{111.1748F,astro-ph/9809085v1,astro-ph/0201143v2,astro-ph/0602326v1,1002.3701v1} over $10\,252\deg^2$ divided in two patches on the sky and cover a redshift range of $z = 0.2 - 0.75$. The final BOSS DR12 analysis splits this redshift range in three overlapping redshift bins defined by $0.2 < z < 0.5$, $0.4 < z < 0.6$ and $0.5 < z < 0.75$ with the effective redshifts $z_{\rm eff} = 0.38$, $0.51$ and $0.61$. In this analysis we will ignore the middle redshift bin, since it is highly correlated with the other two redshift bins and does not add much additional information.

We include three different incompleteness weights to account for shortcomings of the BOSS dataset (see~\citealt{1203.6499v3} and~\citealt{1312.4877v2} for details): a redshift failure weight, $w_{\rm rf}$, a fibre collision weight, $w_{\rm fc}$ and a systematics weight, $w_{\rm sys}$, which is a combination of a stellar density weight and a seeing condition weight. Each galaxy is thus counted as 
\begin{equation}
w_c = (w_{\rm rf} + w_{\rm fc} - 1)w_{\rm sys}.
\end{equation} 
More details about these weights and their effect on the DR12 sample can be found in~\citet{Ross2016:1607.03145v1}. 

\section{BOSS measurements and uncertainties}
\label{sec:measurement}

The power spectrum measurements used in this paper make use of the FFT based estimator~\citep{Bianchi2015:1505.05341v2, Scoccimarro2015:1506.02729v2} and are discussed in more detail in~\citet{Beutler2016:1607.03150v1} and~\citet{Beutler2016:1607.03149v1}.
Here we will summarise these measurements but refer to the above mentioned references for more details.

The first three non-zero power spectrum multipoles can be calculated as~\citep{Feldman1993:astro-ph/9304022v1}
\begin{align}
P_{0}(\vc{k}) &= \frac{1}{2A}\left[F_{0}(\vc{k})F_{0}^*(\vc{k})  - S\right],\\
P_{2}(\vc{k}) &= \frac{5}{4A}F_{0}(\vc{k})\left[3F_2^*(\vc{k}) - F_0^*(\vc{k})\right],\\
P_{4}(\vc{k}) &= \frac{9}{16A}F_{0}(\vc{k})\left[35F_4^*(\vc{k}) - 30F_2^*(\vc{k}) + F_0^*(\vc{k})\right],
\end{align}
where the shot noise and the normalisation are given by 
\begin{align}
S &= (1 + \alpha)\int d^3x n_g(x)w^2_{\rm FKP}(x)\\
A &= \int d^3x n_g(x)w_{\rm FKP}(x)
\end{align}
with $\alpha$ being the ratio between the number of galaxies and randoms. The Fourier-space density moments are given by
\begin{align}
F_{0}(\vc{k}) &= A_0(\vc{k}),\\
\begin{split}
F_{2}(\vc{k}) &= \frac{1}{k^2}\bigg[k_x^2 B_{xx} + k_y^2 B_{yy} + k_z^2B_{zz}\\
&+2\bigg(k_xk_y B_{xy} + k_xk_zB_{xz} + k_yk_zB_{yz}\bigg)\bigg],
\end{split}\\
\begin{split}
F_{4}(\vc{k}) &= \frac{1}{k^4}\bigg[k_x^4 C_{xxx} + k_y^4 C_{yyy} + k_z^4C_{zzz}\\
&+4\bigg(k_x^3k_y C_{xxy} + k^3_xk_zC_{xxz} + k^3_yk_xC_{yyx})\\
&+ k_y^3k_z C_{yyz} + k^3_zk_xC_{zzx} + k^3_zk_yC_{zzy})\bigg)\\
&+6\bigg( k_x^2k_y^2 C_{xyy} + k^2_xk^2_zC_{xzz} + k_y^2k_z^2C_{yzz})\bigg)\\
&+12k_xk_yk_z\left(k_xC_{xyz} + k_yC_{yxz} + k_zC_{zxy}\right)\bigg].
\end{split}
\end{align}
Following~\citet{Bianchi2015:1505.05341v2} and~\citet{Scoccimarro2015:1506.02729v2} we can write
\begin{align}
A_0(\vc{k}) &= \int d\vc{r} D(\vc{r})e^{i\vc{k}\cdot \vc{r}},
\label{eq:ps_eq1}\\
B_{xy}(\vc{k}) &= \int d\vc{r} \frac{r_xr_y}{|\vc{r}|^2}D(\vc{r})e^{i\vc{k}\cdot \vc{r}},
\label{eq:ps_eq2}\\
C_{xyz}(\vc{k}) &= \int d\vc{r} \frac{r^2_xr_yr_z}{|\vc{r}|^4}D(\vc{r})e^{i\vc{k}\cdot \vc{r}},
\label{eq:ps_eq3}
\end{align}
where $D(\vc{r})$ is the galaxy overdensity field. The three equation above can be calculated using FFTs.

\subsection{Covariance matrix}

To derive a covariance matrix for the power spectrum multipoles we use $2048$\footnote{To be precise we have $2048$ mocks for the SGC and $2045$ mocks for the NGC.} MultiDark-Patchy mock catalogues~\citep{Kitaura2015:1509.06400v3}. These mock catalogues have been calibrated to a $N$-body based reference sample using approximate gravity solvers and analytical-statistical biasing models. The reference catalogue is extracted from one of the BigMultiDark simulations~\citep{Klypin2014:1411.4001v2}, which used $3\,840^3$ particles on a volume of ($2.5h^{-1}$Gpc)$^3$ assuming a $\Lambda$CDM cosmology with $\Omega_M = 0.307115$, $\Omega_b = 0.048206$, $\sigma_8 = 0.8288$, $n_s = 0.9611$, and a Hubble constant of $H_0 = 67.77\,$km/s/Mpc.

\subsection{Window function}
\label{sec:win}

Before comparing any model to the power spectrum measurement we convolve it with the survey window function using the technique discussed in section 4 of~\citet{Beutler2016:1607.03150v1}, which is based on~\citet{Wilson2015:1511.07799v2}.
The technique applies the following steps to turn a power spectrum model without any window function effect into the required convolved power spectrum including the survey window function:
\begin{enumerate}
\item Calculate the model power spectrum multipoles and Fourier-transform them to obtain the correlation function multipoles $\xi_{L}^{\rm model}(s)$.
\item Calculate the ``convolved'' correlation function multipoles $\hat{\xi}^{\rm model}_{\ell}(s)$ by multiplying the correlation function with the window function multipoles. 
\item Conduct 1D FFTs to transform the convolved correlation function multipoles back into Fourier space to obtain the convolved power spectrum multipoles, $\hat{P}^{\rm model}_{\ell}(k)$. This result becomes our model to be compared with the observed power spectrum multipoles.
\end{enumerate}
For more details about the implementation we refer to~\citet{Beutler2016:1607.03150v1}.

\section{Power spectrum model}
\label{sec:model}

\begin{figure*}
\begin{center}
\epsfig{file=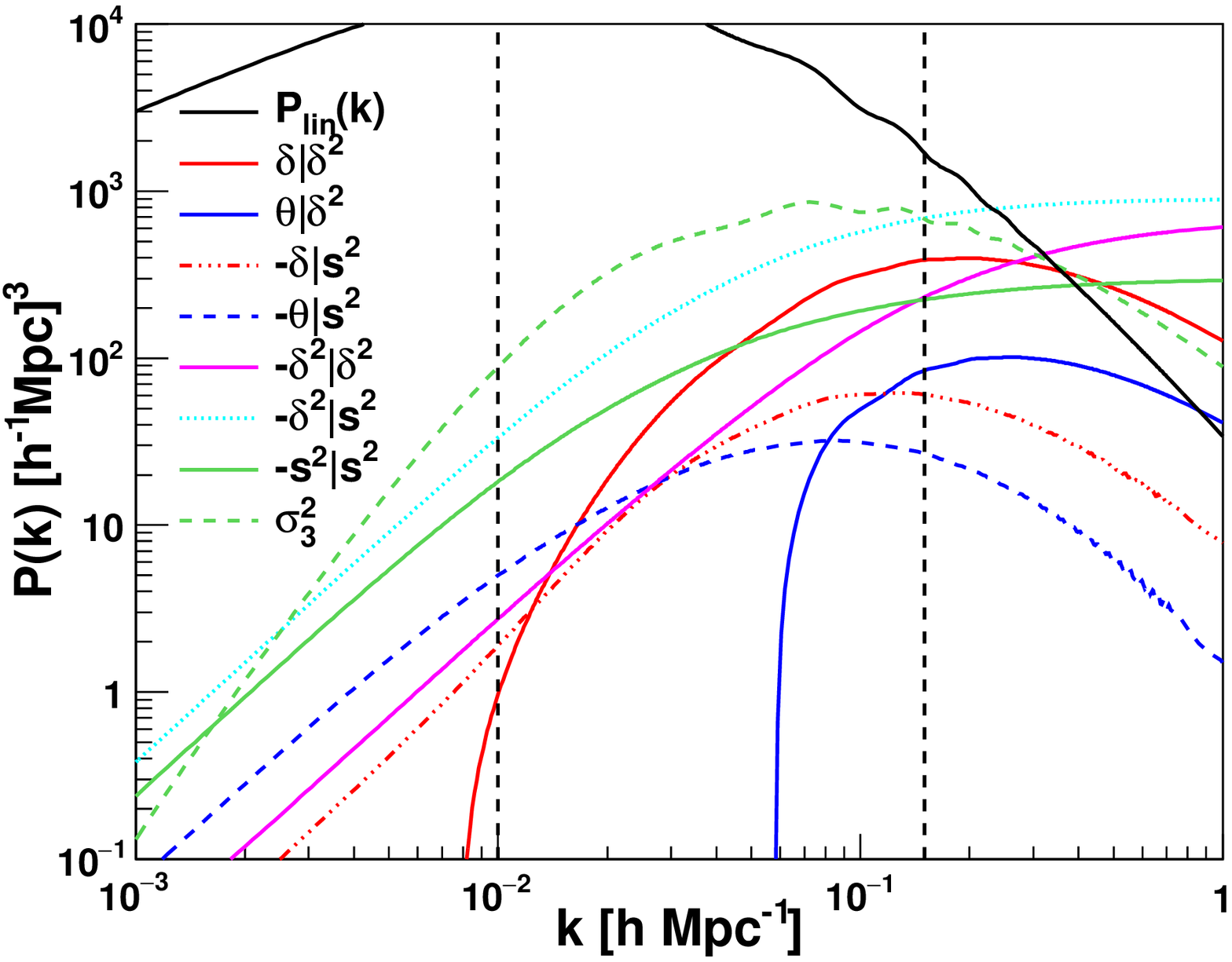,width=5.8cm}
\epsfig{file=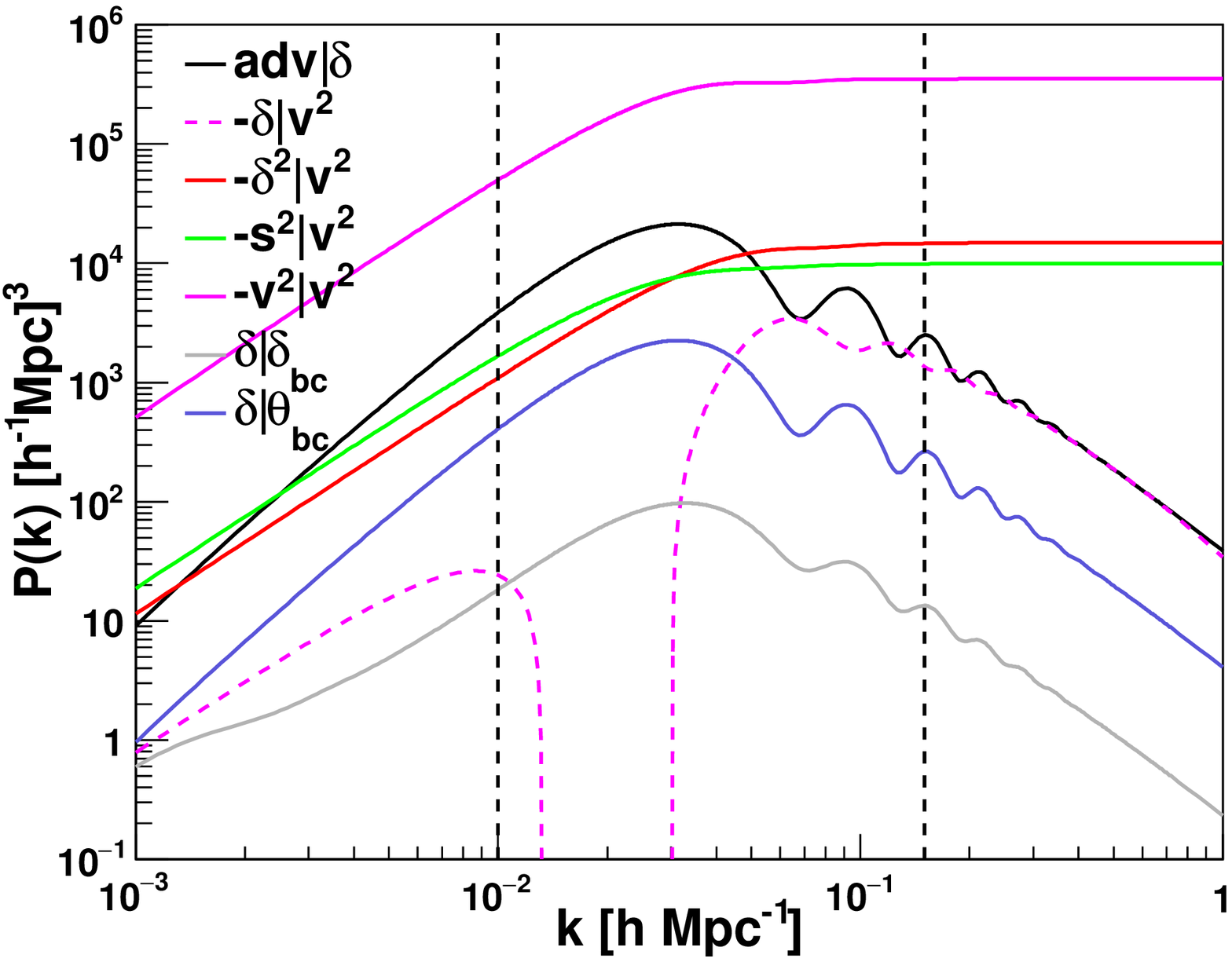,width=5.8cm}
\epsfig{file=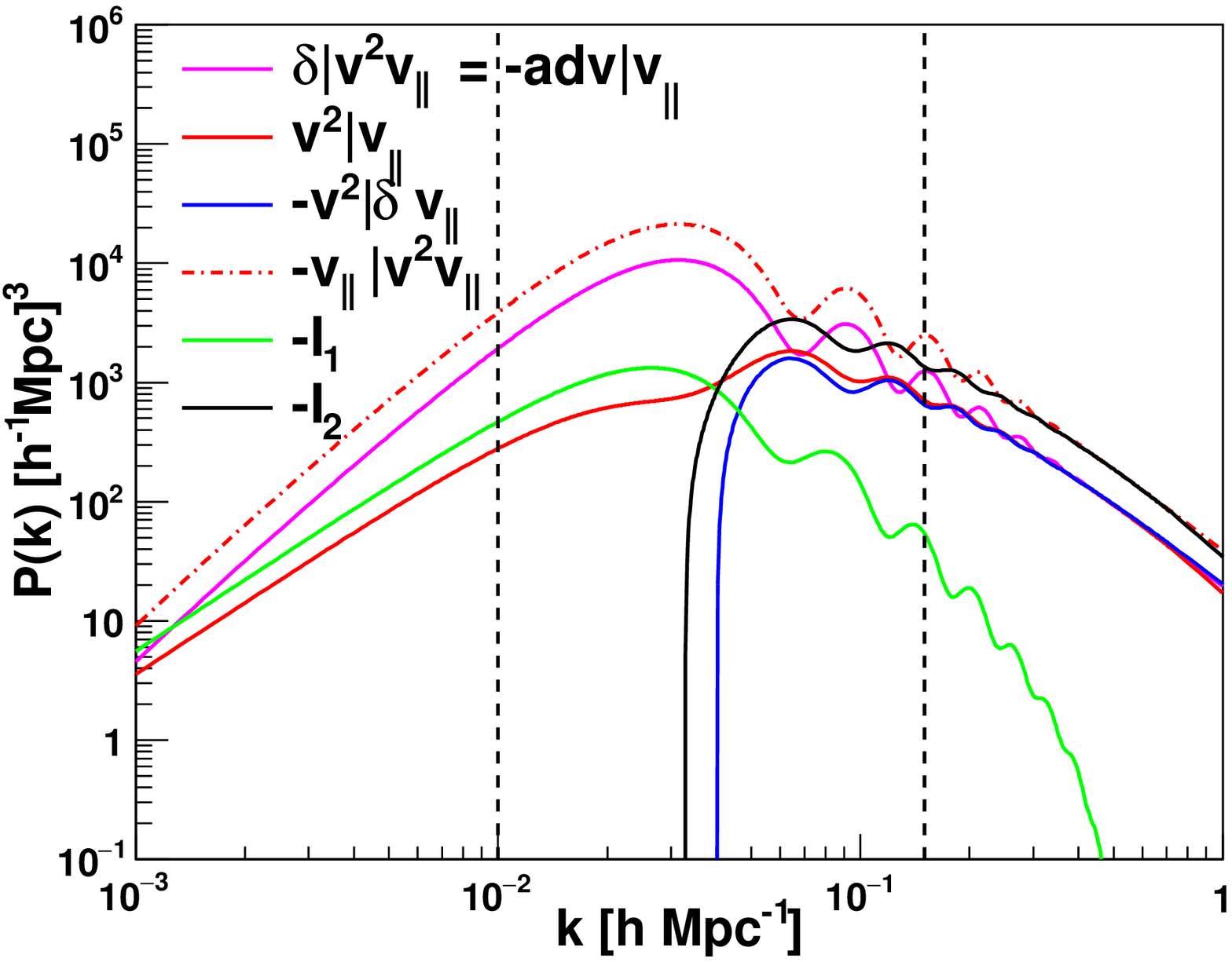,width=5.8cm}
\caption{Comparison of the different perturbative terms used in our power spectrum model (see eq.~\ref{eq:psmodel} and appendix~\ref{app:PTterms}). Left: comparison of the density and velocity terms, middle: comparison of the correlations between the density field and the relative velocity field, right: correlation between the relative velocity field and the velocities. The fitting results presented in this paper make use of the scales between the two dashed lines.}
\label{fig:ps_cmp}
\end{center}
\end{figure*}

\begin{figure*}
\begin{center}
\epsfig{file=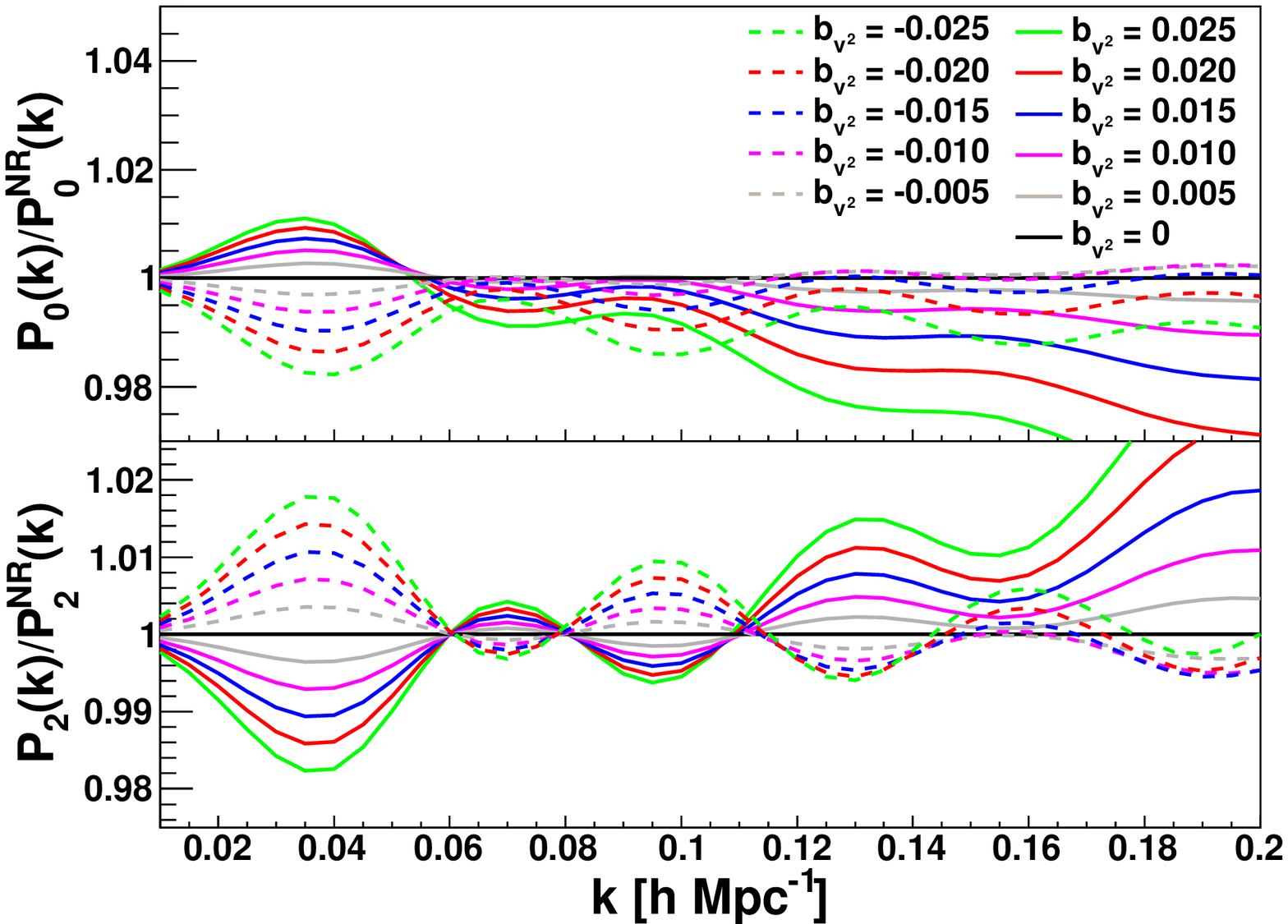,width=8.8cm}
\epsfig{file=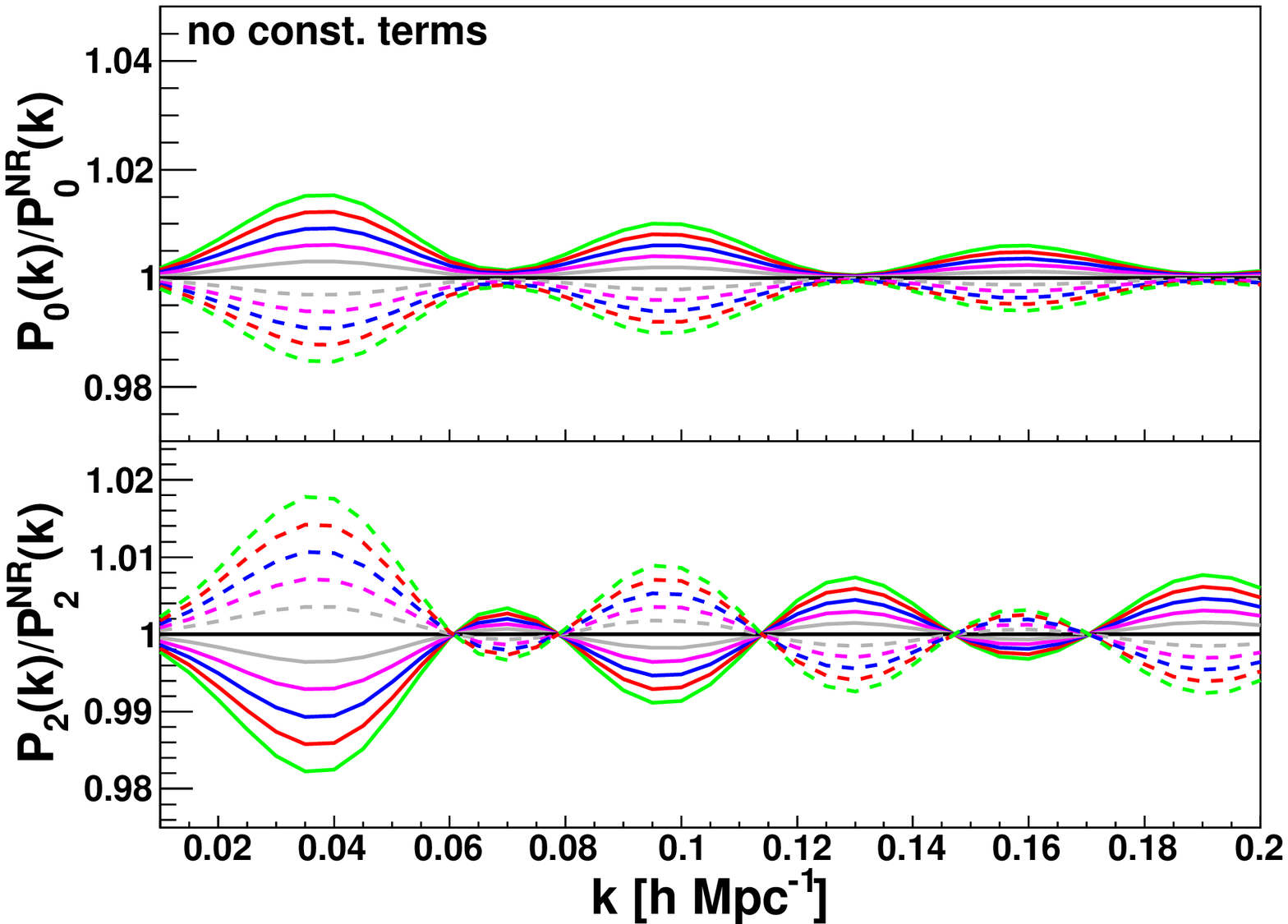,width=8.8cm}
\caption{This plot shows the effect of the $b_{v^2}$ parameter to the power spectrum monopole (top) and quadrupole (bottom). $P_{\ell}^{\rm NR}(k)$ is the power spectrum multipole with all relative velocity parameters set to zero. All other parameters are fixed. The plot on the right excludes the terms $P_{v^2v^2}$, $P_{\delta^2v^2}$ and $P_{s^2v^2}$, in which case $b_{v^2}$ does not have any effect on the amplitude but purely changes the oscillation pattern.}
\label{fig:cmodel}
\end{center}
\end{figure*}

The power spectrum model we employ in this paper is an extension of the model used in~\citet{Beutler2014:1312.4611v2, Beutler2016:1607.03150v1} and builds upon the work of~\citet{1006.0699v1},~\citet{0902.0991v1} and~\citet{1405.1447v4}. Here we extend this model by including the relative velocity terms following the approach of~\citet{1105.3732v1} and~\citet{1510.03554v2} with the addition of redshift-space distortion terms, which describe the couplings of the density field with the velocity divergence field. We also include the linear terms $P_{\delta|\delta_{\rm bc}}(k)$ and $P_{\delta|\theta_{\rm bc}}(k)$ as discussed in~\citet{Schmidt2016:1602.09059v2}.

We define the galaxy density field as 
\begin{equation}
\begin{split}
\delta^s_g(x) &= b_1\delta_m(x) + \frac{1}{2}b_2\left[\delta_m^2(x) - \langle \delta_m^2\rangle\right] + \frac{1}{2}b_s\left[s^2(x) - \langle s^2\rangle\right] + \dots\\
&+ b_{v^2}\left[v^2_{\rm bc}(x) - \langle v_{\rm bc}^2\rangle\right]\\
&+ b^{\rm bc}_{\delta}\left[\delta_b(x) - \delta_c(x)\right] + b^{\rm bc}_{\theta}\left[\theta_b(x) - \theta_c(x)\right] + \dots,\\
\end{split}
\end{equation}
where $\delta_m(x)$ is the matter density field, $v_{\rm bc}(x)$ is the relative velocity field, $s(x)$ is the tidal tensor field, $\delta_{\rm bc}(x)$ is the relative density field between baryons and cold dark matter and $\theta_{\rm bc}(x)$ is the relative velocity divergence field. The power spectrum for the density field above is
\begin{equation}
\begin{split}
P_g(k,\mu) &= P_{\rm g, NL}(k,\mu)+ b_{v^2}\Big[b_1P_{\delta|v^2}(k) +  b_2P_{\delta^2|v^2}(k) \\
&+ b_sP_{s^2|v^2}(k) + b_{v^2}P_{v^2|v^2}(k)\Big]\\
&+ b_1b_{v^2}P_{\rm adv|\delta}(k) + 2b_1b^{\rm bc}_{\delta}P_{\delta|\delta_{\rm bc}} + 2b_1b^{\rm bc}_{\theta}P_{\delta|\theta_{\rm bc}}\\
&- 2f\mu^2\Bigg[b_{v^2}\left(b_1P_{\delta|v^2v_{\parallel}}(k) + P_{\rm adv|v_{\parallel}}(k)\right)\\
&-b^{\rm bc}_{\theta}P_{\delta|\theta_{\rm bc}} + b^{\rm bc}_{\delta}P_{\delta|\delta_{\rm bc}}\\
&+ b_{v^2}\left(P_{v^2|v_{\parallel}}(k) + P_{v^2|\delta v_{\parallel}}(k)\right)\Bigg]\\
& + f^2\mu^4b_{v^2}P_{v_{\parallel}|v^2v_{\parallel}}(k)\\
&- f^2\mu^2b_{v^2}\left[I_{1}(k) + \mu^2I_{2}(k)\right],
\end{split}
\label{eq:psmodel}
\end{equation}
where we ignored the $b^{\rm bc, 2}_{\theta}$ and $b^{\rm bc, 2}_{\delta}$ terms, which in our case are expected to be about one order of magnitude smaller compared to the linear terms~\citep{Schmidt2016:1602.09059v2}. All the different terms in the equation above are defined in appendix~\ref{app:PTterms}. The first term, $P_{\rm g, NL}$, describes the linear and nonlinear terms connecting the real-space matter density field with the redshift-space galaxy density field and is given by
\begin{equation}
\begin{split}
P_{\rm g, NL}(k,\mu) &= \exp\left\{-(fk\mu\sigma_v)^2\right\}\big[P_{{\rm g},\delta\delta}(k)\\
&\;\;\;\; + 2f\mu^2P_{{\rm g},\delta\theta}(k) + f^2\mu^4P_{\theta\theta}(k)\\
&\;\;\;\; + b_1^3A(k,\mu,\beta) + b_1^4B(k,\mu,\beta)\big], 
\label{eq:taruya}
\end{split}
\end{equation}
with
\begin{align}
\begin{split}
P_{{\rm g},\delta\delta}(k) &= b_1^2P_{\delta\delta}(k) + b_2b_1 P_{b2,\delta}(k) 
+ b_{s2}b_1P_{bs2,\delta}(k)\\
& + 2b_{\rm 3nl}b_1\sigma_3^2(k)P^{\rm lin}_{\rm m}(k) + b^2_2P_{b22}(k)\\
& + b_2b_{s2}P_{b2s2}(k) + b^2_{s2}P_{bs22}(k) + N,
\label{eq:40}
\end{split}\\
\begin{split}
P_{{\rm g},\delta\theta}(k) &= b_1P_{\delta\theta}(k) + b_2P_{b2,\theta}(k) 
+ b_{s2}P_{bs2,\theta}(k)\\
& + b_{\rm 3nl}\sigma_3^2(k)P^{\rm lin}_{\rm m}(k).
\label{eq:41}
\end{split}
\end{align}
The terms A and B in eq.~\ref{eq:taruya} account for coupling between the density field and the velocity field~\citep{Taruya2010:1006.0699v1}, $\sigma_v$ is a free parameter describing the velocity dispersion on quasi-linear scales and $N$ is another free parameter used to marginalise over any constant non-Poisson shot noise. This is the base redshift-space model of~\citet{McDonald2009:0902.0991v1},~\citet{Taruya2010:1006.0699v1} and~\citet{Saito2014:1405.1447v4}, which has been tested extensively in~\citet{Beutler2014:1312.4611v2, Beutler2016:1607.03150v1}. In this paper we focus on the relative velocity extensions to this model.
The dominant terms in eq.~\ref{eq:psmodel}, with respect to the relative velocity effects, are
\begin{align}
P_{\rm adv|\delta}(k) &= \frac{4}{3}T_{v}(k)kP_{\rm lin}(k)\int \frac{k\,dk}{2\pi^2}T_{v}(k)P_{\rm lin}(k)\\
P_{\delta|v^2}(k) &= 4\int \frac{d^3\vc{q}}{(2\pi)^3}P^{\rm lin}_{\rm m}(q)P^{\rm lin}_{\rm m}(k-q)\\
&\times F_2(\vc{q},\vc{k-q})G_u(\vc{q},\vc{k-q}) \frac{\vc{q}\cdot(\vc{k-q})}{\vc{q}(\vc{k-q})}\\
P_{\delta|\delta_{\rm bc}}(k) &= T_{\rm bc}(k)P_{\rm lin}(k)\\
P_{\delta|\theta_{\rm bc}}(k) &= \frac{\sigma_{vbc}}{H_0}T_{v}(k)kP_{\rm lin}(k)
\end{align}
with the kernels
\begin{align}
F_2(\vc{k_1},\vc{k_2}) &= \frac{5}{7} + \frac{\vc{k_1}\cdot \vc{k_2}}{2}\left(\frac{1}{k_1^2} + \frac{1}{k_2^2}\right) + \frac{2}{7}\left(\frac{\vc{k_1}\cdot \vc{k_2}}{k_1k_2}\right)^2,\\
G_u(\vc{k_1},\vc{k_2}) &= -T_v(k_1)T_{v}(k_2)
\end{align}
and the velocity transfer function
\begin{equation}
T_{v}(k) \propto \frac{T_{v_b}(k) - T_{v,cdm}(k)}{T_m(k)},
\label{eq:Tv}
\end{equation}
where $T_{v_b}$ and $T_{v,cdm}$ are the velocity transfer functions of baryons and cold dark matter, respectively.
The matter transfer function equivalent is defined as 
\begin{equation}
T_{\rm bc}(k) = \frac{T_{b}(k) - T_{cdm}(k)}{T_m(k)}.
\end{equation}
The normalisation for the velocity transfer function is given by the square root of 
\begin{equation}
\sigma^2_{vbc} = \int\frac{k^2\,dk}{2\pi^2}T_{v}^2(k)P_{\rm lin}(k),
\label{eq:Tvnorm}
\end{equation}
which is dimensionless, since $T_v$ as defined in eq.~\ref{eq:Tv} is dimensionless. Note that the advection term and the relative velocity divergence term are related by $P_{\rm adv|\delta}(k) = AP_{\delta|\theta_{\rm bc}}(k)$ with 
\begin{equation}
A = \frac{4H_0}{3\sigma_{vbc}}\int \frac{kdk}{2\pi^2}T_v(k)P(k),
\end{equation} 
where we use $H_0^{-1} = 2997\,$Mpc and $\sigma_{vbc} = 1.64\times10^{-6}$, resulting in $A = 1820$ at $z = 0.38$ and $A = 2044$ at $z = 0.61$. While $P_{\delta|\delta_{\rm bc}}(k)$ constrains the bias parameter $b^{\rm bc}_{\delta}$ and $P_{\delta|\theta_{\rm bc}}(k)$ constrains $b^{\rm bc}_{\theta}$, the relative velocity bias $b_{v^2}$ is constrained by the sum of $P_{\rm adv|\delta}(k)$ and $P_{\delta|v^2}(k)$.

We follow the nomenclature of~\citet{1510.03554v2} meaning that our velocity bias $b_{v^2}$ is a factor of $3$ times smaller compared to~\citet{1308.1401v2}. A list of all terms in eq.~\ref{eq:psmodel} is given in appendix~\ref{app:PTterms} and included in Figure~\ref{fig:ps_cmp}. The Figure clearly highlights the oscillations present in some of the relative velocity terms. These oscillations are the main reason for our study, since these oscillations are out-of-phase with the baryon acoustic oscillations and therefore represent a potential bias when measuring the BAO scale.

In our fits we do not vary $b_s$ and $b_{\rm 3nl}$ freely, but fix them to 
\begin{align}
b_{s} &= -\frac{4}{7}(b_1 - 1),\\
b_{\rm 3nl} &= \frac{32}{315}(b_1 - 1),
\label{eq:bias_relations}
\end{align}
which is in good agreement with what is observed in simulations~\citep{1405.1447v4} and can be motivated from theory~\citep{Baldauf2012:1201.4827v1, Chan2012:1201.3614v2, Saito2014:1405.1447v4}. See also~\citealt{Desjacques2016:1611.09787v1} for a recent review on large scale galaxy bias.

\begin{figure}
\begin{center}
\epsfig{file=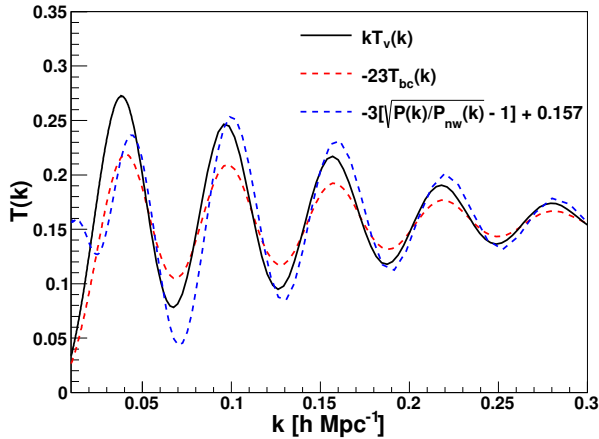,width=8.8cm}
\caption{This plot compares the BAO signature in the transfer function $T_v$, which is underlying the advection term $P_{\delta|\theta_{\rm bc}}$, and $T_{bc}$, which underlies the term $P_{\delta|\delta_{\rm bc}}$. We also include the $P/P_{\rm nw}$ term, which describes the linear BAO. The $P/P_{\rm nw}(k)$ and $T_{\rm bc}$ terms are scaled, to put all functions on the same scale. The different phases of these oscillations are the reason why the relative velocity effect is a potential systematic for BAO measurements.}
\label{fig:tk}
\end{center}
\end{figure}

\subsection{Discussion of the power spectrum model}

The relative velocity density field $\delta_{\rm bc}$ describes the variation in the cold dark matter to baryon ratio given the fact that baryons and cold dark matter start off with different initial conditions after decoupling. The relative velocity divergence $\theta_{\rm bc}$ captures the same effect in the velocity field. 
The term $P_{\delta|\delta_{\rm bc}}(k)$ corresponds to correlations between variations of the baryon to cold dark matter ratio and the overall matter density field and $P_{\delta|\theta_{\rm bc}}(k)$ corresponds to correlations of the relative velocity divergence fields with the overall matter density field.  While the first term is expected to be of order $1$, the second term is expected to be of order $\approx 6.8\left[(1+z)H_0\right]^{-1}(b_1 - 1)$~\citep{Schmidt2016:1602.09059v2}. All terms which are proportional to $b_{v^2}$ decay with redshift ($\propto 1/a$).

Our power spectrum model uses the {\tt class}~\citep{Lesgourgues2011:1104.2932v2} transfer function output to calculate the velocity transfer function in eq.~\ref{eq:Tv}. At high redshift the relative velocity transfer function evolves with the scale factor, which does not enter our calculation, since this scaling is removed by our normalisation in eq.~\ref{eq:Tvnorm}. Since we assume that all imprints of the relative velocity effects come from $z > 15$ we use the $z = 15$ transfer function and ignore any low redshift effects.

\section{Test on mock catalogues}
\label{sec:systests}

We first test our power spectrum model on N-body simulations before using the BOSS Mutidark Patchy mock catalogues. 

\begin{figure*}
\begin{center}
\epsfig{file=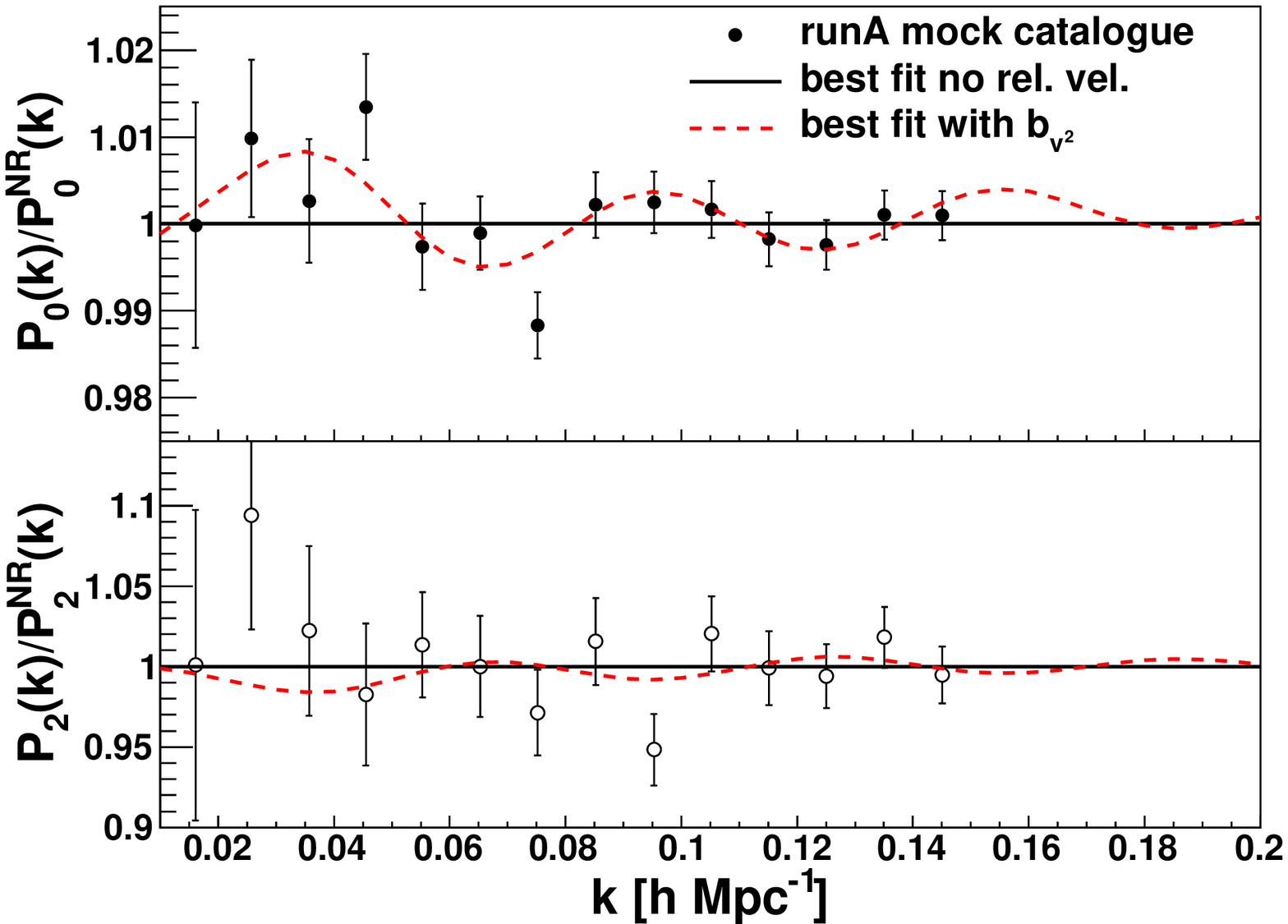,width=8cm}
\epsfig{file=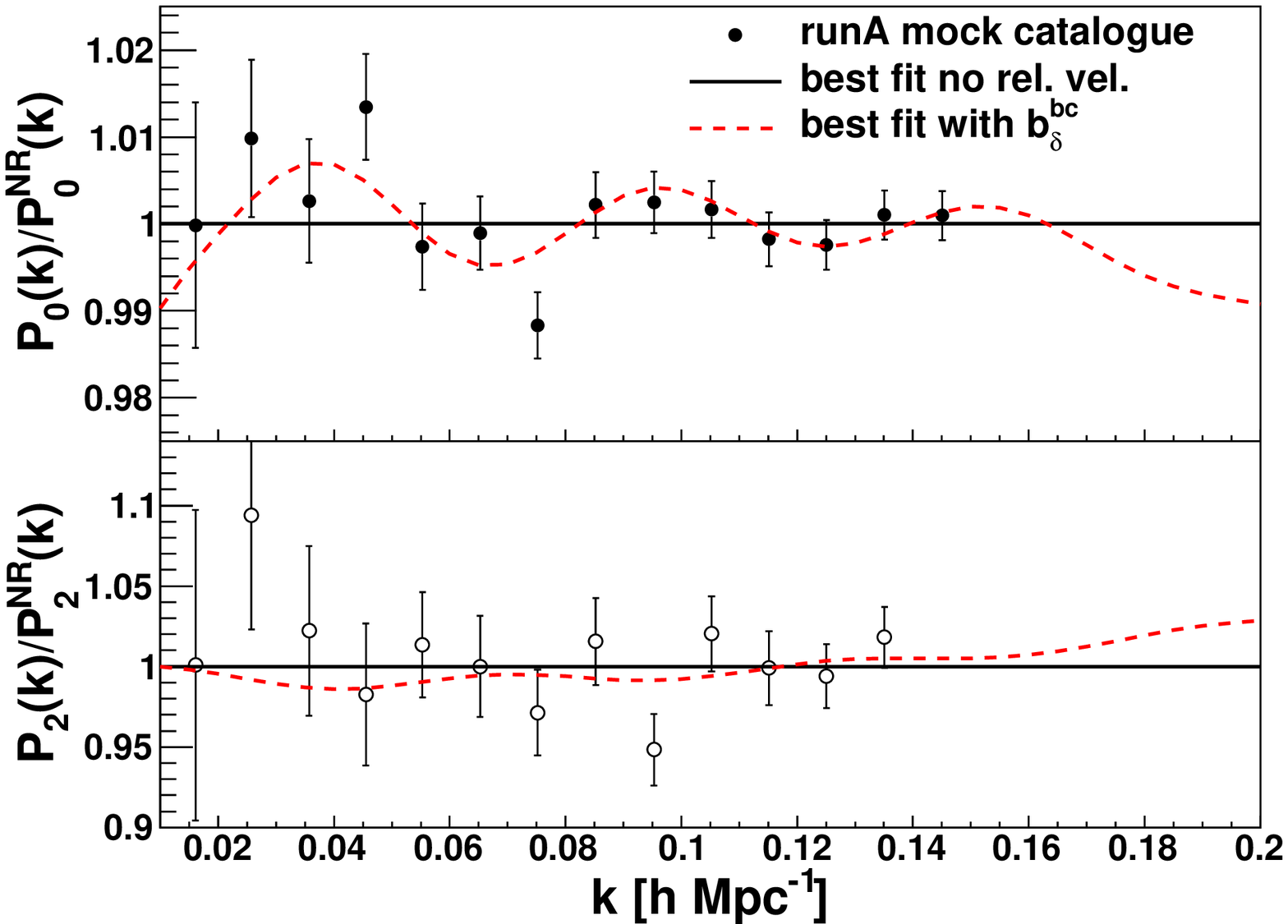,width=8cm}
\caption{These plots compare the best fitting model for the runA simulations setting all relative velocity parameters to zero (black line) with the fit including $b_{v^2}$ (left, red dashed line) and $b^{\rm bc}_{\delta}$ (right, red dashed line). $P^{\rm NR}(k)$ refers to the power spectrum model with all relative velocity parameters set to zero. The solid points show the mean monopole measurements for the 20 runA simulations and the open points show the equivalent for the quadrupole. The $\Delta\chi^2$ between the solid black line and the red dashed line is $20.9 - 16.2 = 4.7$ for $b_{v^2}$ and $20.9 - 15.9 = 5.0$ for $b^{\rm bc}_{\delta}$. This means we have a moderate $2.2\sigma$ significance for a non-zero value for these bias parameters, even though these values are expected to be zero, given that the simulations do not include a relative velocity effect.}
\label{fig:ratio_runA_data}
\end{center}
\end{figure*}

\begin{table*}
\centering
\caption{This table shows the fitting results to the runA, runPB and the Multidark Patchy mock catalogues including the relative velocity parameters $b_{v^2}$, $b^{\rm bc}_{\delta}$ and $b^{\rm bc}_{\theta}$. For these tests we fix the parameters $\alpha_{\perp}$, $\alpha_{\parallel}$ and $f\sigma_8$ to their fiducial values. Note that these tests have been done for each parameter separately meaning the constraints on $b_{v^2}$ assume $b^{\rm bc}_{\delta} = b^{\rm bc}_{\theta} = 0$ etc. The errors refer to the $1\sigma$ and $2\sigma$ (in parentheses) uncertainties. All simulations show consistent results for the three bias parameters, including a systematic shift, which we take into account when fitting the data (see section~\ref{sec:analysis}).}
      \begin{tabular}{lllllllll}
         \hline
         & \multicolumn{2}{c}{only $b_{v^2}$ (runA)} & \multicolumn{2}{c}{only $b_{v^2}$ (runPB)} & \multicolumn{2}{c}{only $b_{v^2}$ (Patchy $z_{1}$)} & \multicolumn{2}{c}{only $b_{v^2}$ (Patchy $z_{3}$)}\\
         \hline
          & max. like. & mean & max. like. & mean & max. like. & mean & max. like. & mean\\
          $\alpha_{\perp}$  & $ 1 $ & $ 1 $ & $ 1$ & $ 1 $ & $ 1$ & $ 1 $ & $ 1$ & $ 1 $\\
          $\alpha_{\parallel}$  & $ 1 $ & $ 1 $ & $ 1 $ & $ 1 $ & $ 1$ & $ 1 $ & $ 1$ & $ 1 $\\
          $f\sigma_8$  & $ 0.455 $ & $ 0.455 $ & $ 0.472 $ & $ 0.472 $ & $ 0.484 $ & $ 0.484 $ & $ 0.478 $ & $ 0.478 $\\
          $b_{v^2}[10^{-3}]$  & $ 21.9 $ & $ 22.2 \pm 6.8 (\pm14) $ & $19$ & $20\pm11(\pm21)$ & $ 29.1 $ & $ 29.8\pm 5.0(\pm9.6) $ & $ 27.6 $ & $ 27.0^{+6.2}_{-7.9}(^{+19}_{-22}) $\\
          $b^{\rm bc}_{\delta}$  & $ -3.6 $ & $ -3.5 \pm 1.1(\pm2.1) $ & $ -2.2 $ & $ -2.3 \pm 1.5(\pm3.0) $ & $ -4.96 $ & $ -4.78 \pm 0.78(\pm1.6) $ & $ -3.44 $ & $ -3.47 \pm 0.66(\pm1.3) $ \\
          $b^{\rm bc}_{\theta}$  & $ 142 $ & $ 147 \pm 51(^{+170}_{-98}) $ & $ 82 $ & $ 77 \pm 63(\pm120) $ & $ 187.2 $ & $ 187.0 \pm 6.8(\pm9.6) $ & $ 191.9 $ & $ 192.5\pm 6.5(\pm9.4) $ \\
         \hline
      \end{tabular}
       \label{tab:sys}
\end{table*}

\subsection{Test on N-body simulations}

To test our fitting technique we use two different sets of N-body simulations, designated as runA and runPB. The runA simulations are $20$ halo catalogues of size $[1500\hMpc]^3$ with $1500^3$ particles using the fiducial cosmology of $\Omega_m = 0.274$, $\Omega_{\Lambda} = 0.726$, $n_s = 0.95$, $\Omega_b = 0.0457$, $H_0 = 70\,$km\,s$^{-1}$Mpc$^{-1}$ $f\sigma_8(z=0.55) = 0.455$ and $r_s(z_d) = 104.503\hMpc$. The runPB simulations are $10$ galaxy catalogues of size $[1380\hMpc]^3$ with $\Omega_m = 0.292$, $\Omega_{\Lambda} = 0.708$, $n_s = 0.965$, $\Omega_b = 0.0462$, $H_0 = 69\,$km\,s$^{-1}$Mpc$^{-1}$, $f\sigma_8(z=0.55)=0.472$ and $r_s(z_d) = 102.3477\hMpc$. The runPB simulations make use of a CMASS-like halo occupation distribution (HOD) model to populate dark matter halos with galaxies (see~\citealt{Reid:2014iaa} for details). The fundamental modes for these simulations are $2\pi/[1500\,\text{Mpc}/h] = 0.0042h/$Mpc for runA and $2\pi/[1380\,\text{Mpc}/h] = 0.0046h/$Mpc for runPB, which is below the $k_{\rm min}=0.01h/$Mpc used in our fits.

We measure the power spectrum monopole, quadrupole and hexadecapole and fit these measurements with the model discussed in the last section. Given that we are working with periodic boxes, we can ignore window function effects for now. The results are summarised in Table~\ref{tab:runA} and \ref{tab:runPB}. For these tests we fix the cosmological parameters ($\alpha_{\parallel}$, $\alpha_{\perp}$ and $f\sigma_8$) to their fiducial values.

\subsection{Fits to runA simulations}

A table summarising the fitting results for the runA simulations is included in the appendix (Table~\ref{tab:runA}). When varying the individual relative velocity parameters, we see significant biases (at the level of $3\sigma$) in all three relative velocity parameters, while there are no biases if $b_{v^2}$ and $b_{\delta}$ are varied simultaneously. However, degeneracies between the parameters increase the uncertainties by factors of $3$ and $1.3$ for $b_{v^2}$ and $b^{\rm bc}_{\delta}$, respectively compared to the fits where each is varied individually.

In Figure~\ref{fig:ratio_runA_data} we compare the best fitting models with and without $b_{v^2}$ and $b^{\rm bc}_{\delta}$. While the bias in both parameters is only on the three sigma level, it seems to be driven by small scales.

\subsection{Fits to runPB simulations}

A table summarising the fitting results for the runPB simulations is included in the appendix (Table~\ref{tab:runPB}). The fits to runPB are consistent with what we saw for the runA simulations, even though the significance of the detected bias in the relative velocity terms is now $< 2\sigma$ due to the larger uncertainties in the runPB mocks. 

\subsection{Tests on the Multidark Patchy mock catalogues}

In Table~\ref{tab:patchy_z1_red} and~\ref{tab:patchy_z3_red} we included the results when fitting the mean of the Multidark Patchy power spectra for the high and low redshift bins. These fits now include the window function treatment described in section~\ref{sec:win}. The results are consistent with the runA and runPB simulations, meaning we detect a shift in all three relative velocity parameters.

\subsection{Summary: Model tests with simulations}

We summarised the results for the three different bias parameters from the three mock catalogues in Table~\ref{tab:sys}.

Given that none of our mock catalogues includes the relative velocity effect, we expect all relative velocity parameters to be consistent with zero. However, we detected shifts in the relative velocity parameters, which are consistent in all three sets of mock catalogues. We investigated these biases further, by (1) only using the monopole, (2) replacing the Multidark Patchy covariance matrix with a linear Gaussian covariance matrix, (3) using the real-space power spectrum instead of the one in redshift-space, (4) varying $b_{s^2}$ and $b_{3nl}$ freely instead of fixing them by the relations in eq.~\ref{eq:bias_relations} and (5) introducing the leading scale dependent bias term $2b_1R^2k^2P_{\rm lin}(k)$ to eq.~\ref{eq:40}~\citep{Okumura2015:1506.05814v2}. None of these changes to the model was able to explain the biases we measure. We therefore conclude that these biases represent a shortcoming of our model.

The detected shifts are of the order of $1\sigma$ when compared to the measurement uncertainties on these parameters we report in section~\ref{sec:analysis}. Therefore, they are not negligible and need to be taken into account when analysing the BOSS power spectrum.

Using the fitting results of Table~\ref{tab:sys} we can quantify the systematic shifts in the parameters of interest. The uncertainty weighted mean for all three simulations is $b_{v^2} = 0.0265\pm0.0033$, $b^{\rm bc}_{\delta} = -3.79\pm0.44$ and $b^{\rm bc}_{\theta} = 187.2\pm 4.7$. 

For the case where we have $b_{v^2}$ and $b^{\rm bc}_{\delta}$ as free parameters we also have to account for their correlation. We found mean shifts from the truth of $0.036$ in $b_{v^2}$ and $1.5$ in $b^{\rm bc}_{\delta}$. The correlation between these two values is $77\%$ and the covariance matrix is 
\begin{equation}
C = \left(\begin{matrix}0.033  &  3.629\\
 3.629 &  676.6\end{matrix}\right)\times 10^{-3},
\end{equation}
where the top left corner corresponds to the $b_{v^2}$ auto-correlation and the bottom right corner corresponds to the $b^{\rm bc}_{\delta}$ auto-correlation. When fitting the data we correct the best fitting values by these systematic shifts and include the error on these values in the error budget.

\section{BOSS DR12 analysis}
\label{sec:analysis}

We are now fitting the power spectrum multipoles using the model of section~\ref{sec:model} including the relative velocity terms. \citet{Schmidt2016:1602.09059v2} suggests that the dominant relative velocity contribution is given by $b^{\rm bc}_{\delta}$ followed by $b_{v^2}$, while the contribution by $b^{\rm bc}_{\theta}$ should be quite small. We fit each relative velocity parameter in turn but also consider the two parameter extension with the two dominant terms $b^{\rm bc}_{\delta}$ and $b_{v^2}$. Our fits include the monopole and quadrupole in the range $0.01 < k < 0.15h^{-1}$Mpc and the hexadecapole with $0.01 < k < 0.10h^{-1}$Mpc. The systematic uncertainties on the relative velocity parameters have been quantified in section~\ref{sec:systests} and we will correct our best fitting values by the observed systematic shift. We also include the error on the systematic shift in our error budget. We note that the systematic shifts we found in our tests on mock catalogues are $< 2\sigma$ of the BOSS measurement uncertainties and the error on the systematic shift is not contributing significantly to our error budget.

As discussed in~\citet{Beutler2016:1607.03150v1} we use separate nuisance parameters for the NGC and SGC, given small differences in their selection, which affect the bias parameters. We ignored the middle redshift bin of BOSS DR12, which has been used in other studies of this dataset, since it is strongly correlated with the other two redshift bins and does not provide much additional information.

We summarise our fitting results for the two redshift bins and the three relative velocity parameters in Table~\ref{tab:combined_z1_relvel_boss} and \ref{tab:combined_z3_relvel_boss}. The BOSS DR12 data does not support a detection of any of the three relative velocity parameters. The reduced $\chi^2$ for the high redshift bin is slightly below $1$, while for the high redshift bin this quantity is slightly above $1$ consistent with the findings of~\citet{Beutler2016:1607.03150v1}. The p-values provided in brackets indicate that these deviations from unity are not significant. 

Combining the high and low redshift bins we find the following limits on the three relative velocity parameters: $b_{v^2} = 0.012 \pm 0.015 (\pm 0.031)$, $b^{\rm bc}_{\delta} = -1.0 \pm 2.5 (\pm 6.2)$ and $b^{\rm bc}_{\theta} = -114 \pm 55 (\pm 175)$ with $68\%$ ($95\%$) confidence levels.

If we treat the relative velocity effect as a pure suppression of star formation in regions where the relative velocity exceeds the virial velocity of halos, we can apply a prior of $b_{v^2} < 0$~\citep{Dalal2010:1009.4704v1}. This improves our constraints on $b_{v^2}$ to $|b_{v^2}| < 0.007 (< 0.018)$ ($68\%$ and $95\%$ confidence levels).

\begin{table*}
   \begin{center}
      \caption{Fits to the BOSS DR12 combined sample power spectrum multipoles in the low redshift bin $0.2 < z < 0.5$. The fit includes the monopole and quadrupole between $0.01 < k < 0.15 h^{-1}$Mpc and the hexadecapole between $0.01 < k < 0.10 h^{-1}$Mpc. All errors in this Table are the marginalised $68\%$ confidence levels, except of the error on the relative velocity parameters $b_{v^2}$, $b^{\rm bc}_{\delta}$ and $b^{\rm bc}_{\theta}$, where we show both, the $68\%$ and $95\%$ confidence levels. We show fits including each relative velocity parameter in turn meaning column $2$ and $3$ show the fits with $b_{v^2}$ as a free parameter assuming $b^{\rm bc}_{\delta} = b^{\rm bc}_{\theta} = 0$ etc. The relative velocity parameters are corrected by the bias we detected in the mock catalogues ($b_{v^2} = 0.0265\pm0.0033$, $b^{\rm bc}_{\delta} = -3.79\pm0.44$ and $b^{\rm bc}_{\theta} = 187.2\pm 4.7$, $[b_{v^2}, b^{\rm bc}_{\delta}] = [0.036, 1.5]$), where the last term in the square brackets includes the correlation between $b_{v^2}$ and $b^{\rm bc}_{\delta}$ used for the combined fits in column $8$ and $9$. These fits show no evidence for a significant detection of any of the relative velocity parameters.}
      \begin{tabular}{lllllllll}
         \hline
         \multicolumn{9}{c}{Fit to the data}\\
 & \multicolumn{2}{c}{+ $b_{v^2}$} & \multicolumn{2}{c}{+ $b^{\rm bc}_{\delta}$} & \multicolumn{2}{c}{+ $b^{\rm bc}_{\theta}$} & \multicolumn{2}{c}{+ $b_{v^{2}} + b^{\rm bc}_{\delta}$}\\
         \hline
 & max. like. & mean & max. like. & mean & max. like. & mean & max. like. & mean\\
          $\alpha_{\perp}$  & $ 1.000 $ & $ 1.002 \pm 0.032 $ & $ 1.008 $ & $ 1.009 \pm 0.029 $ & $ 1.007 $ & $ 1.012 \pm 0.029 $ & $ 1.004 $ & $ 1.007 \pm 0.030 $ \\
          $\alpha_{\parallel}$  & $ 0.999 $ & $ 1.004 \pm 0.043 $ & $ 1.004 $ & $ 1.007 \pm 0.040 $ & $ 1.003 $ & $ 1.007 \pm 0.043 $ & $ 1.004 $ & $ 1.007 \pm 0.039 $ \\
          $f\sigma_8$  & $ 0.480 $ & $ 0.481 \pm 0.060 $ & $ 0.480 $ & $ 0.485 \pm 0.062 $ & $ 0.476 $ & $ 0.477 \pm 0.061 $ & $ 0.465 $ & $ 0.466 \pm 0.063 $ \\
          $b_{v^2}[10^{-3}]$  & $ 14 $ & $ 19 \pm 21 (\pm 44 )$ & $ 0 $ & $ 0 $ & $ 0 $ & $ 0 $ & $ 24 $ & $ 24 ^{+ 18 }_{- 14 } (^{+ 54 }_{- 34 }) $ \\
          $b_{\delta}$  & $ 0 $ & $ 0 $ & $ 1.4 $ & $ 1.4 \pm 4.3 (^{+ 9.0 }_{- 12.0 }) $ & $ 0 $ & $ 0 $ & $ 6.2 $ & $ 6.4 \pm 6.3 (\pm 13.0 )$ \\
          $b_{\theta}$  & $ 0 $ & $ 0 $ & $ 0 $ & $ 0 $ & $ -71 $ & $ -67 \pm 81 (\pm 270 )$ & $ 0 $ & $ 0 $ \\
\hline
          $b_{1}^{\rm NGC}\sigma_8$  & $ 1.324 $ & $ 1.316 \pm 0.047 $ & $ 1.346 $ & $ 1.348 \pm 0.052 $ & $ 1.33 $ & $ 1.335 \pm 0.052 $ & $ 1.358 $ & $ 1.351 \pm 0.049 $ \\
          $b_{1}^{\rm SGC}\sigma_8$  & $ 1.325 $ & $ 1.322 \pm 0.058 $ & $ 1.340 $ & $ 1.340 \pm 0.060 $ & $ 1.330 $ & $ 1.333 \pm 0.060 $ & $ 1.371 $ & $ 1.362 \pm 0.054 $ \\
          $b^{\rm NGC}_2\sigma_8$  & $ 1.33 $ & $ 1.31 \pm 0.76 $ & $ 1.20 $ & $ 1.32 \pm 0.71 $ & $ 0.56 $ & $ 0.77 \pm 0.76 $ & $ 1.58 $ & $ 1.28 \pm 0.83 $ \\
          $b^{\rm SGC}_2\sigma_8$  & $ 0.7 $ & $ 0.9 \pm 1.0 $ & $ 0.52 $ & $ 0.67 \pm 0.89 $ & $ 0.3 $ & $ 0.6 \pm 1.0 $ & $ 1.24 $ & $ 1.22 \pm 0.95 $ \\
          N$^{\rm NGC}$  & $ -1000 $ & $ -300 \pm 1700 $ & $ -2600 $ & $ -2700 ^{+ 1500 }_{- 1200 } $ & $ -1100 $ & $ -1600 ^{+ 2300 }_{- 1600 } $ & $ -200 $ & $ 300 ^{+ 1500 }_{- 1200 } $ \\
          N$^{\rm SGC}$  & $ -1000 $ & $ -600 \pm 2000 $ & $ -1700 $ & $ -2100 ^{+ 2700 }_{- 1900 } $ & $ -900 $ & $ -1700 ^{+ 3500 }_{- 2300 } $ & $ -900.0 $ & $ -400 \pm 1600 $ \\
          $\sigma^{\rm NGC}_v$  & $ 5.85 $ & $ 5.79 \pm 0.64 $ & $ 5.80 $ & $ 5.80 \pm 0.66 $ & $ 5.63 $ & $ 5.63 \pm 0.70 $ & $ 5.93 $ & $ 5.88 \pm 0.69 $ \\
          $\sigma^{\rm SGC}_v$  & $ 6.52 $ & $ 6.56 \pm 0.85 $ & $ 6.44 $ & $ 6.50 \pm 0.81 $ & $ 6.35 $ & $ 6.36 \pm 0.81 $ & $ 6.70 $ & $ 6.66 \pm 0.80 $ \\
         \hline
         $\frac{\chi^2}{d.o.f.}$ 
& \multicolumn{2}{c}{$\frac{ 79.4 }{ 74 - 12 }=1.28\;(p=0.067)$} & \multicolumn{2}{c}{$\frac{ 80.5 }{ 74 - 12 }=1.30\;(p=0.057)$} & \multicolumn{2}{c}{$\frac{ 80.8 }{ 74 - 12 }=1.30\;(p=0.055)$} & \multicolumn{2}{c}{$\frac{ 78.3 }{ 74 - 13 }=1.28\;(p=0.067)$} \\
         \hline
      \end{tabular}
      \label{tab:combined_z1_relvel_boss}
   \end{center}
\end{table*}

\begin{table*}
   \begin{center}
      \caption{Fits to the BOSS DR12 combined sample power spectrum multipoles in the high redshift bin $0.5 < z < 0.75$. The fit includes the monopole and quadrupole between $0.01 < k < 0.15 h^{-1}$Mpc and the hexadecapole between $0.01 < k < 0.10 h^{-1}$Mpc. All errors in this Table are the marginalised $68\%$ confidence levels, except of the error on the relative velocity parameters $b_{v^2}$, $b^{\rm bc}_{\delta}$ and $b^{\rm bc}_{\theta}$, where we show both, the $68\%$ and $95\%$ confidence levels. We show fits including each relative velocity parameter in turn meaning column $2$ and $3$ show the fits with $b_{v^2}$ as a free parameter assuming $b^{\rm bc}_{\delta} = b^{\rm bc}_{\theta} = 0$ etc. The relative velocity parameters are corrected by the bias we detected in the mock catalogues ($b_{v^2} = 0.0265\pm0.0033$, $b^{\rm bc}_{\delta} = -3.79\pm0.44$ and $b^{\rm bc}_{\theta} = 187.2\pm 4.7$, $[b_{v^2}, b^{\rm bc}_{\delta}] = [0.036, 1.5]$), where the last term in the square brackets includes the correlation between $b_{v^2}$ and $b^{\rm bc}_{\delta}$ used for the combined fits in column $8$ and $9$. These fits show no evidence for a significant detection of any of the relative velocity parameters.}
      \begin{tabular}{lllllllll}
         \hline
         \multicolumn{9}{c}{Fit to the data}\\
 & \multicolumn{2}{c}{+ $b_{v^2}$} & \multicolumn{2}{c}{+ $b^{\rm bc}_{\delta}$} & \multicolumn{2}{c}{+ $b^{\rm bc}_{\theta}$} & \multicolumn{2}{c}{+ $b_{v^{2}} + b^{\rm bc}_{\delta}$}\\
         \hline
 & max. like. & mean & max. like. & mean & max. like. & mean & max. like. & mean\\
          $\alpha_{\perp}$  & $ 0.973 $ & $ 0.979 \pm 0.028 $ & $ 0.971 $ & $ 0.975 \pm 0.030 $ & $ 0.983 $ & $ 0.987 \pm 0.026 $ & $ 0.972 $ & $ 0.976 \pm 0.032 $ \\
          $\alpha_{\parallel}$  & $ 0.975 $ & $ 0.984 \pm 0.043 $ & $ 0.980 $ & $ 0.987 \pm 0.042 $ & $ 0.978 $ & $ 0.984 \pm 0.043 $ & $ 0.980 $ & $ 0.985 \pm 0.047 $ \\
          $f\sigma_8$  & $ 0.419 $ & $ 0.413 \pm 0.047 $ & $ 0.416 $ & $ 0.409 \pm 0.054 $ & $ 0.425 $ & $ 0.421 \pm 0.048 $ & $ 0.420 $ & $ 0.417 \pm 0.056 $ \\
          $b_{v^2}[10^{-3}]$  & $ 1 $ & $ 4 \pm 21 (\pm 43 )$ & $ 0 $ & $ 0 $ & $ 0 $ & $ 0 $ & $ -56 $ & $ -52 \pm 30 (\pm 58 )$ \\
          $b_{\delta}$  & $ 0 $ & $ 0 $ & $ -2.3 $ & $ -2.3 \pm 3.1 (\pm 7.7 )$ & $ 0 $ & $ 0 $ & $ -10.4 $ & $ -10.8 \pm 3.6 (\pm 8.9 )$ \\
          $b_{\theta}$  & $ 0 $ & $ 0 $ & $ 0 $ & $ 0 $ & $ -152 $ & $ -155 \pm 76 (\pm 230 )$ & $ 0 $ & $ 0 $ \\
\hline
          $b_{1}^{\rm NGC}\sigma_8$  & $ 1.219 $ & $ 1.232 \pm 0.045 $ & $ 1.231 $ & $ 1.238 \pm 0.046 $ & $ 1.163 $ & $ 1.162 \pm 0.057 $ & $ 1.230 $ & $ 1.230 \pm 0.060 $ \\
          $b_{1}^{\rm SGC}\sigma_8$  & $ 1.239 $ & $ 1.243 \pm 0.047 $ & $ 1.227 $ & $ 1.232 \pm 0.050 $ & $ 1.262 $ & $ 1.261 \pm 0.049 $ & $ 1.222 $ & $ 1.219 \pm 0.055 $ \\
          $b^{\rm NGC}_2\sigma_8$  & $ 2.94 $ & $ 2.83 ^{+ 0.49 }_{- 0.61 } $ & $ 0.72 $ & $ 1.18 ^{+ 0.94 }_{- 1.20 } $ & $ -1.26 $ & $ -1.39 ^{+ 0.68 }_{- 0.55 } $ & $ 0.66 $ & $ 0.77 \pm 1.20 $ \\
          $b^{\rm SGC}_2\sigma_8$  & $ 0.81 $ & $ 0.94 \pm 0.79 $ & $ 0.74 $ & $ 0.85 \pm 0.88 $ & $ 0.93 $ & $ 0.90 \pm 0.93 $ & $ 0.68 $ & $ 0.68 \pm 0.70 $ \\
          N$^{\rm NGC}$  & $ 0 $ & $ 0 \pm 1800 $ & $ -1000 $ & $ -1600 ^{+ 2400 }_{- 1100 } $ & $ 4700 $ & $ 5200 \pm 2400 $ & $ -1000 $ & $ -1100 \pm 2700 $ \\
          N$^{\rm SGC}$  & $ -500 $ & $ -300 \pm 1400 $ & $ -1000 $ & $ -1200 ^{+ 1700 }_{- 1200 } $ & $ -1500 $ & $ -1400 ^{+ 2000 }_{- 1400 } $ & $ -1000 $ & $ -700 \pm 1600 $ \\
          $\sigma^{\rm NGC}_v$  & $ 5.33 $ & $ 5.31 \pm 0.75 $ & $ 5.11 $ & $ 5.10 \pm 0.80 $ & $ 4.36 $ & $ 4.3 \pm 1.0 $ & $ 5.06 $ & $ 5.02 \pm 0.83 $ \\
          $\sigma^{\rm SGC}_v$  & $ 4.94 $ & $ 4.94 \pm 0.88 $ & $ 4.79 $ & $ 4.70 \pm 0.91 $ & $ 4.99 $ & $ 4.86 \pm 0.90 $ & $ 4.74 $ & $ 4.66 \pm 0.97 $ \\
         \hline
         $\frac{\chi^2}{d.o.f.}$ 
& \multicolumn{2}{c}{$\frac{ 51.7 }{ 74 - 12 }=0.83\;(p=0.821)$} & \multicolumn{2}{c}{$\frac{ 55.3 }{ 74 - 12 }=0.89\;(p=0.714)$} & \multicolumn{2}{c}{$\frac{ 52.0 }{ 74 - 12 }=0.84\;(p=0.813)$} & \multicolumn{2}{c}{$\frac{ 55.2 }{ 74 - 13 }=0.90\;(p=0.685)$} \\
         \hline
      \end{tabular}
      \label{tab:combined_z3_relvel_boss}
   \end{center}
\end{table*}

\section{Quantifying the potential systematic uncertainties for BAO and RSD}
\label{sec:sysbias}

\begin{figure*}
\begin{center}
\epsfig{file=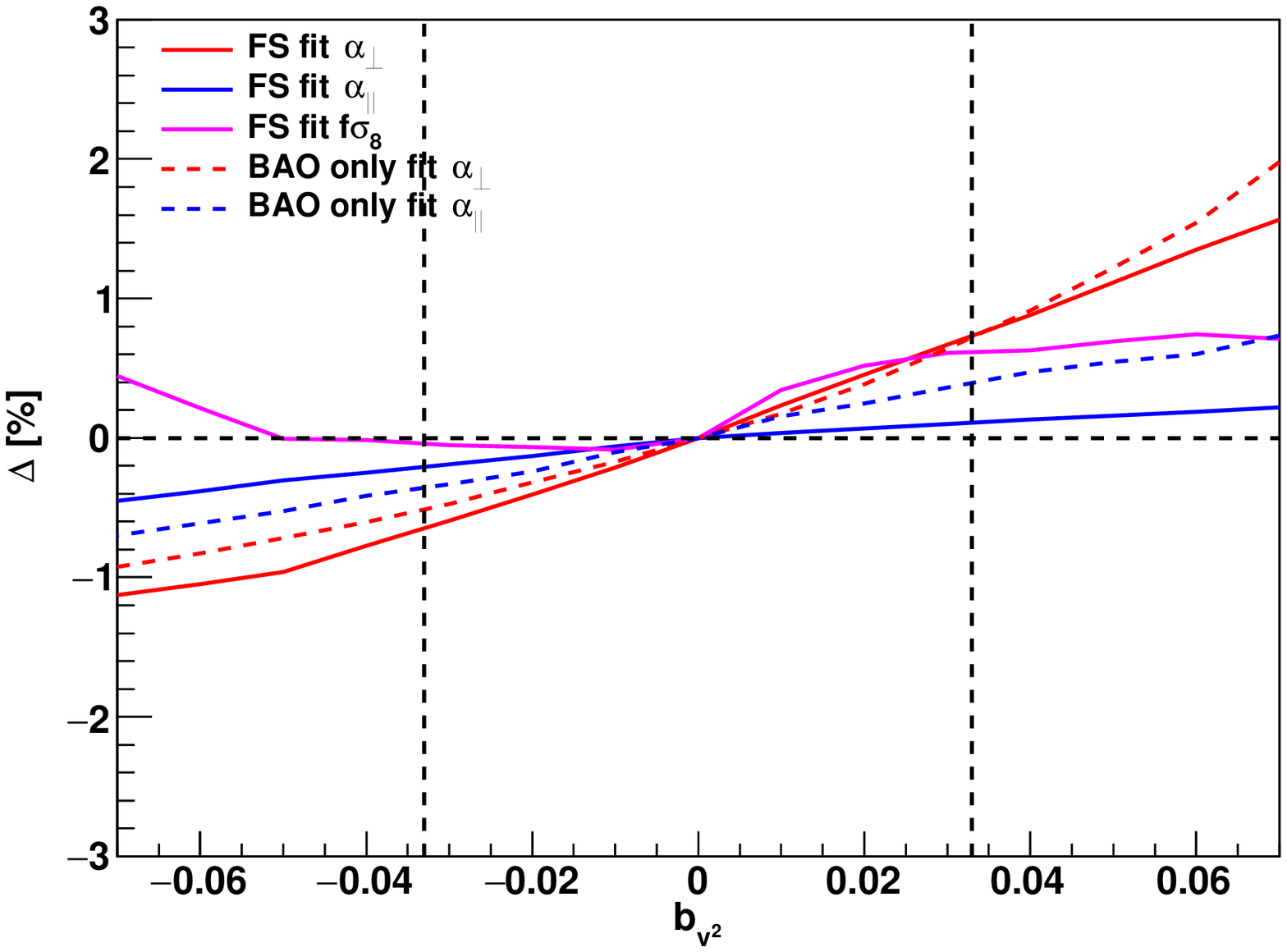,width=8.8cm}
\epsfig{file=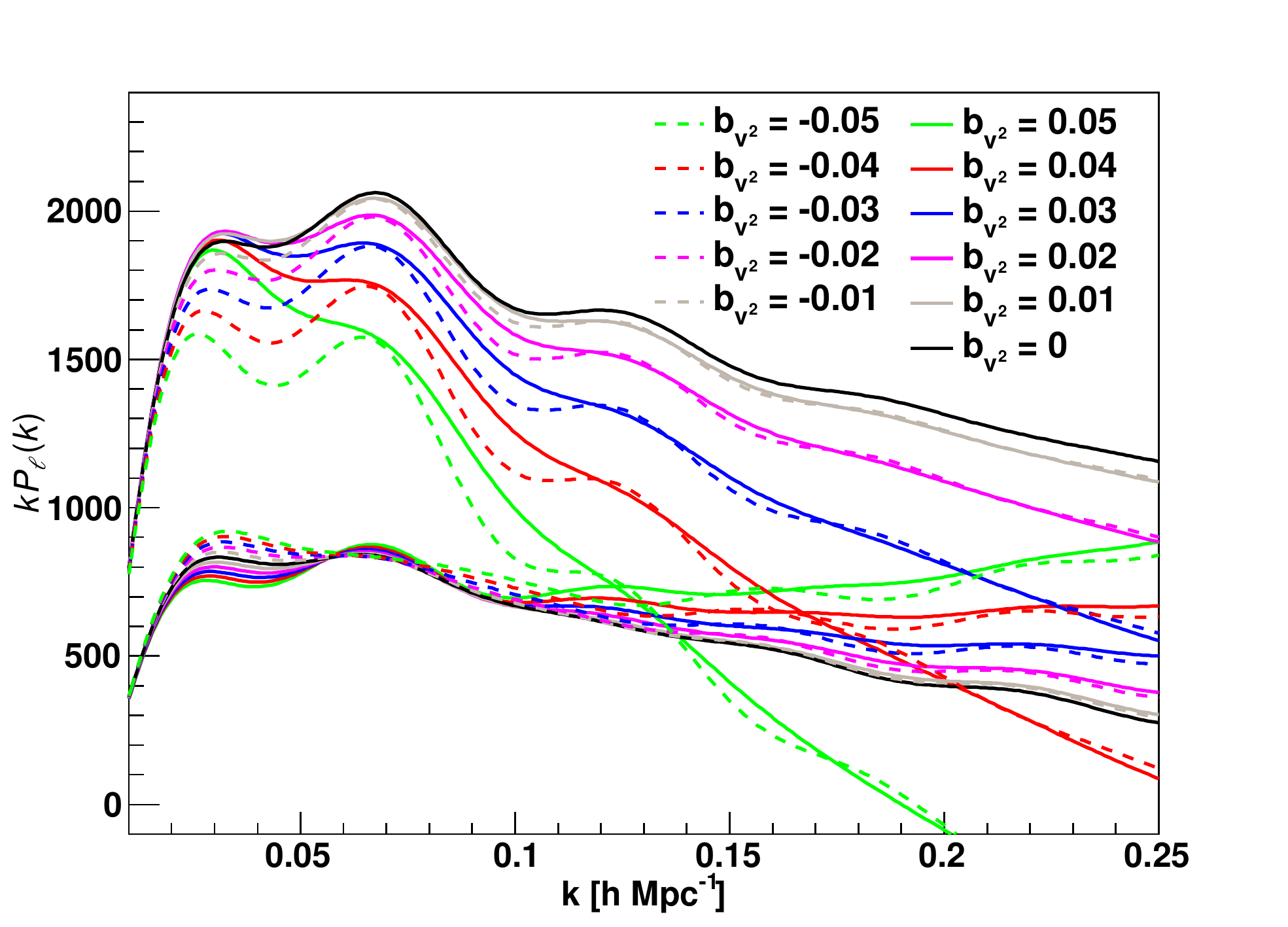,width=8.8cm}\\
\epsfig{file=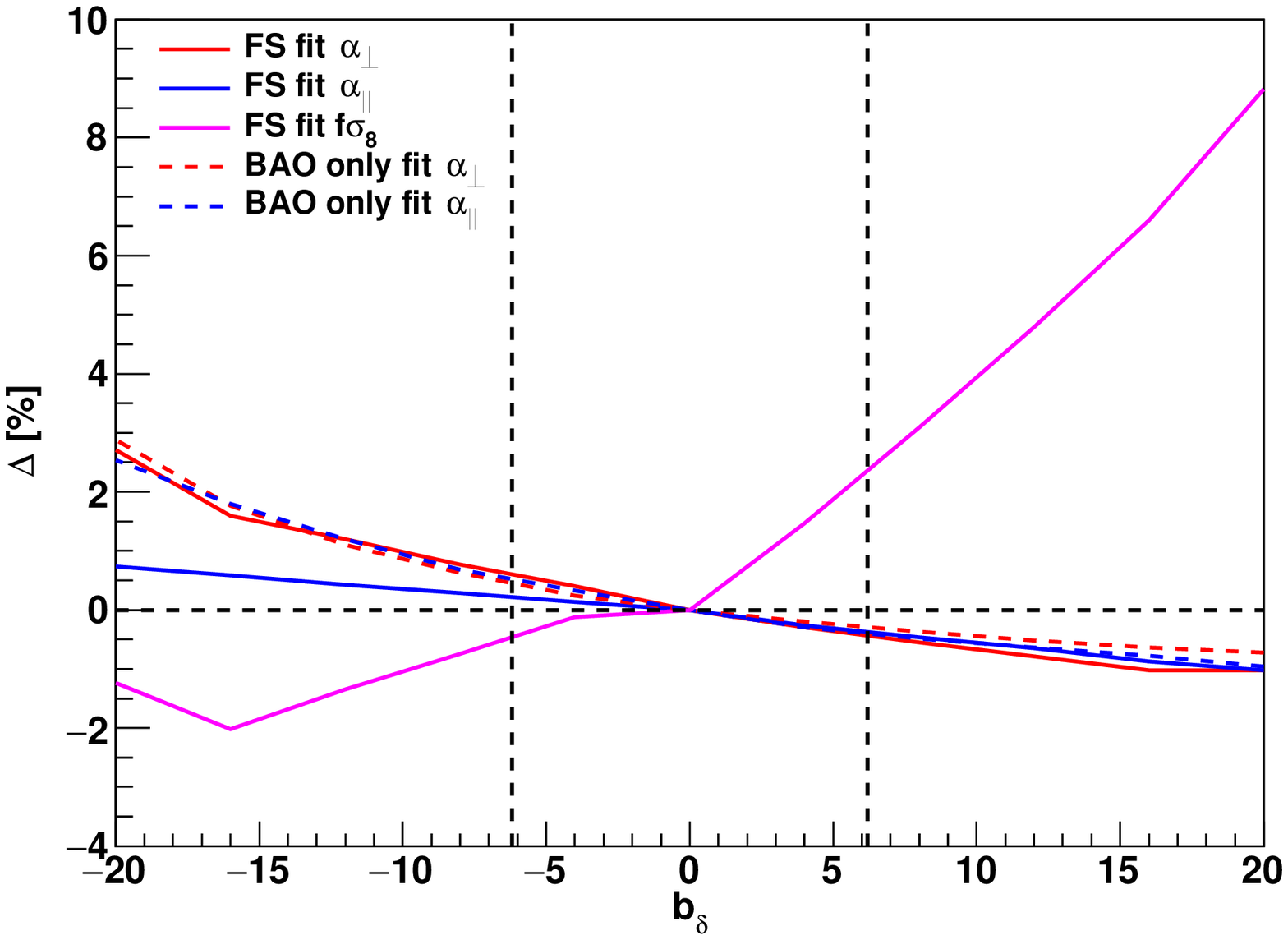,width=8.8cm}
\epsfig{file=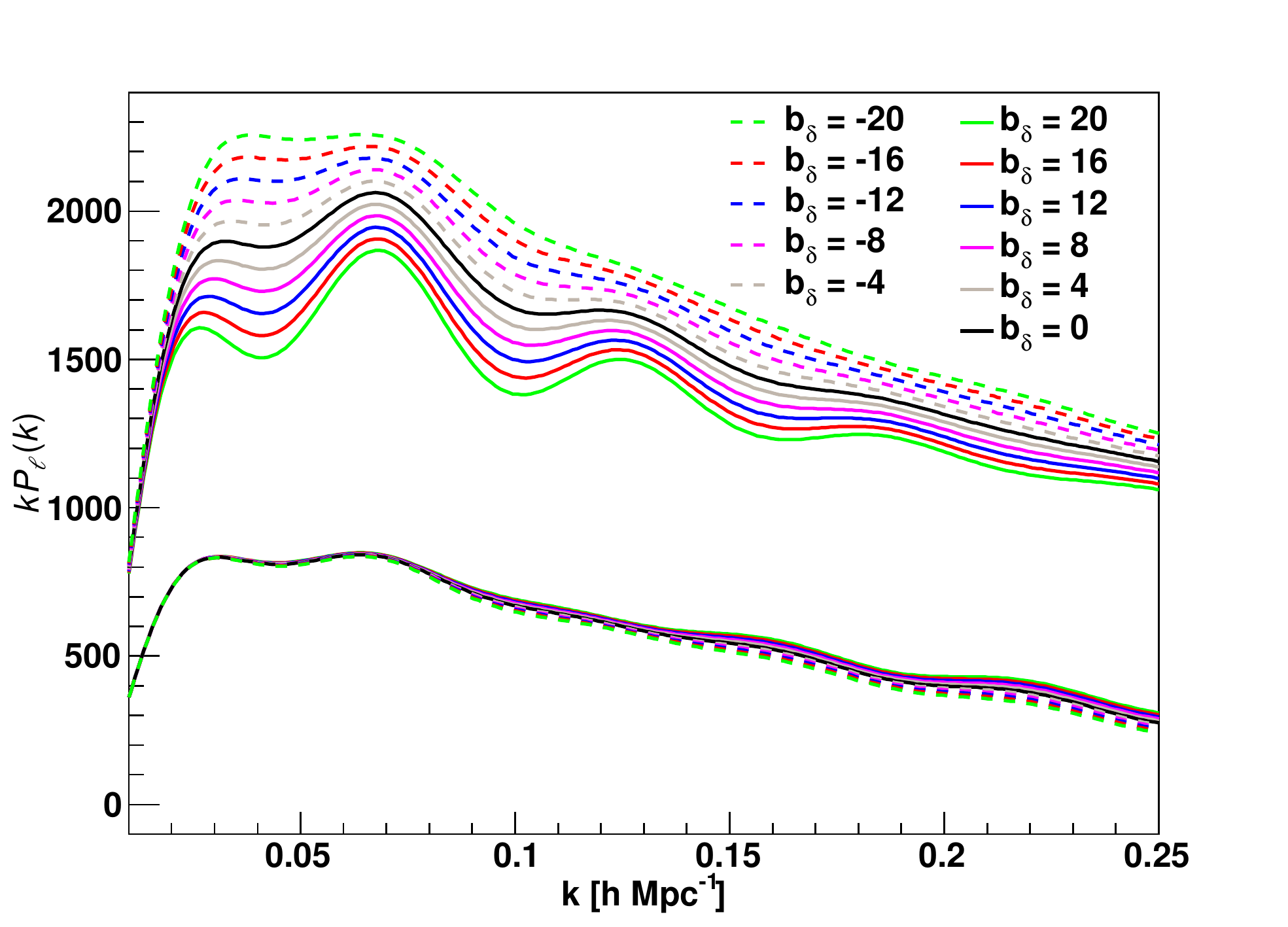,width=8.8cm}\\
\epsfig{file=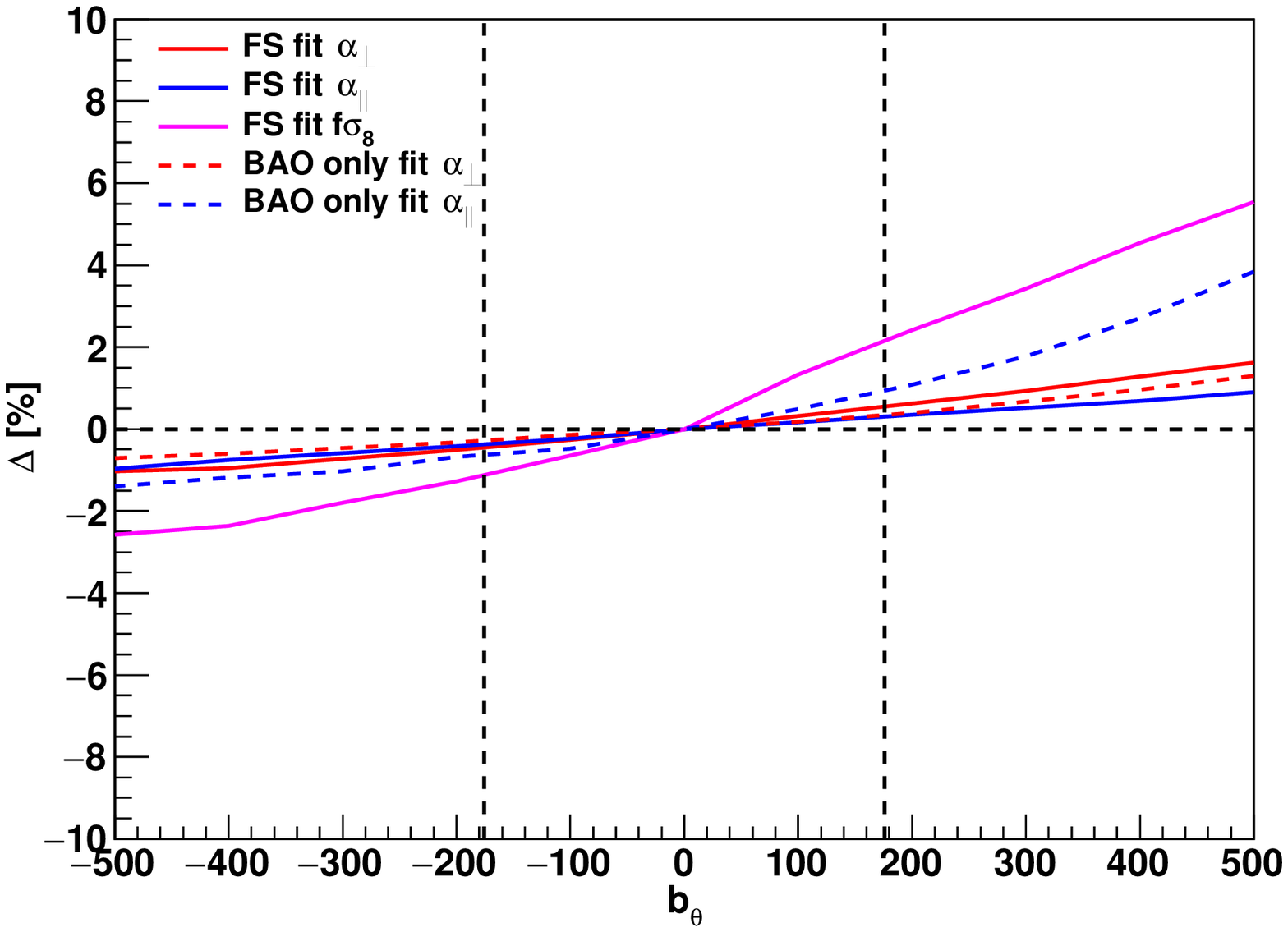,width=8.8cm}
\epsfig{file=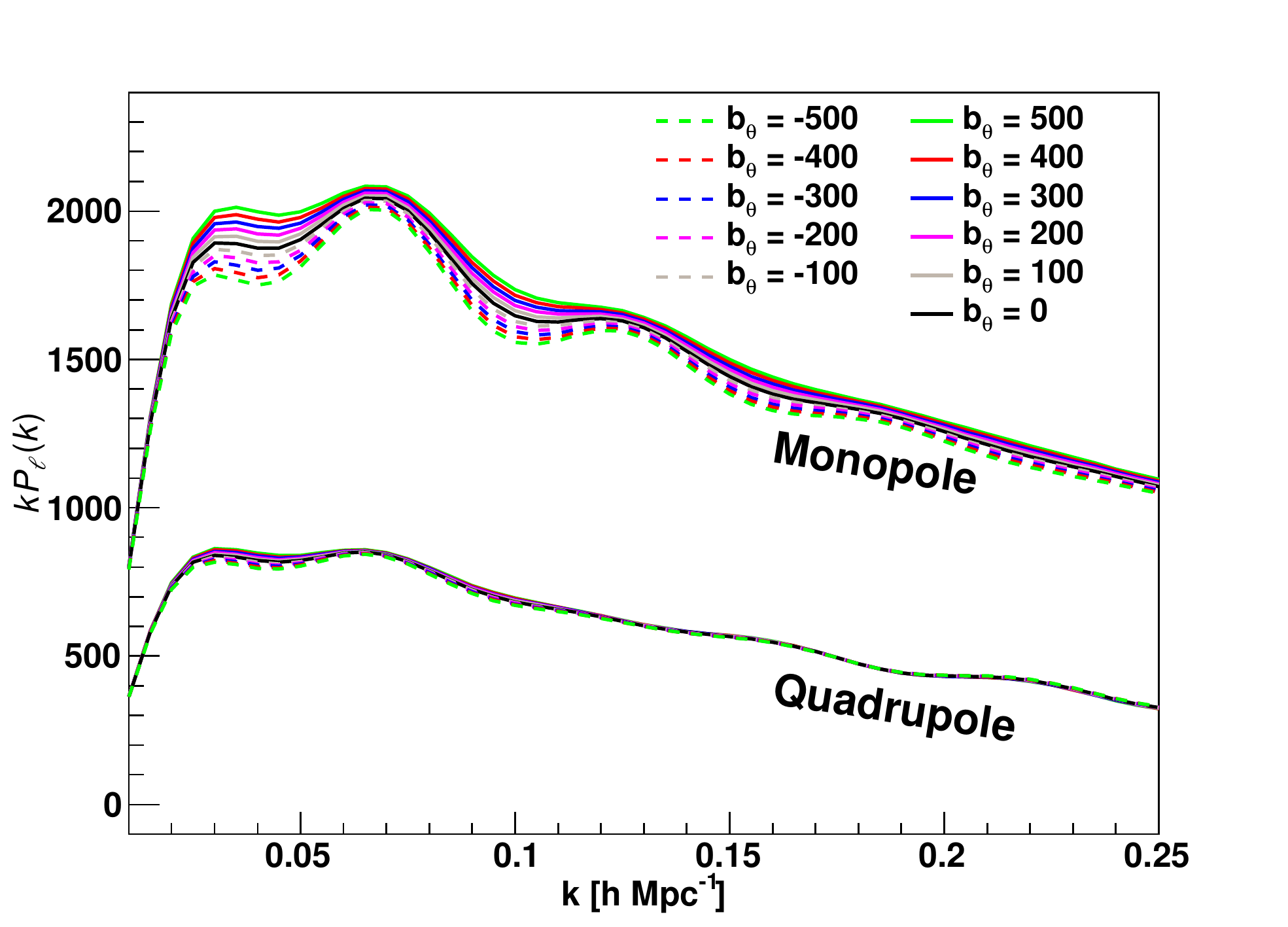,width=8.8cm}
\caption{Here we show the dependence of the shift parameters $\alpha_{\perp}$ and $\alpha_{\parallel}$ as well as the growth of structure parameter $f\sigma_8$ on the three relative velocity parameters (left) and the change in the power spectrum model (right). The solid lines in the plots on the left show the `full shape' (FS) fits using the analysis pipeline of~\citet{Beutler2016:1607.03150v1}, while the dashed lines use the BAO only analysis pipeline of~\citet{Beutler2016:1607.03149v1}. The vertical black dashed lines show the $95\%$ confidence levels for the three relative velocity parameters obtained in this paper (see section~\ref{sec:analysis}).}
\label{fig:shift}
\end{center}
\end{figure*}

Here we want to quantify the potential bias for the anisotropic BAO parameters as well as the RSD parameter, depending on the amplitude of the three relative velocity parameters. To do this we generate power spectrum models as shown in section~\ref{sec:model} and fit these models with the BAO-only fitting pipeline of~\citet{Beutler2016:1607.03149v1} and the `full shape' pipeline of~\citet{Beutler2016:1607.03150v1}. The results are shown in Figure~\ref{fig:shift}. The vertical black dashed lines show the $95\%$ confidence levels from our analysis. 

All three relative velocity parameters are able to shift the BAO scale. The biases are quite different for the two BAO scaling parameters, $\alpha_{\perp}$ and $\alpha_{\parallel}$. The largest shift in $\alpha_{\perp}$ is due to $b_{v^2}$ and reaches $0.8\%$ at $b_{v^2} = 0.031$ (which is the $95\%$ confidence limit we found). The angular BAO scale $\alpha_{\parallel}$ shows $1\%$ shifts due to $b^{\rm bc}_{\theta}$.

We also include the shift in the RSD parameter $f\sigma_8$. Given that $b^{\rm bc}_{\delta}$ and $b^{\rm bc}_{\theta}$ mainly change the monopole to quadrupole ratio, we can see large effects on the RSD parameter of up to $2\%$ in both $b^{\rm bc}_{\delta}$ and $b^{\rm bc}_{\theta}$. Note that the latest measurement from the BOSS survey reported constraints of $1.5\%$ on $D_A(z)$, $2\%$ on $H(z)$ and $9\%$ on $f\sigma_8$\footnote{Here we quote the combined constraints from the two independent redshift bins.}. 

\section{Discussion}
\label{sec:discussion}

In~\citet{Alam2016:1607.03155v1} the potential impact of the relative velocity effect on the BOSS-DR12 BAO measurement has been investigated using a configuration-space model following~\citet{1510.03554v2}. The potential shift in the isotropic BAO scale ($\alpha$) has been limited to $0.3\sigma$, which is consistent with our results for $b_{v^2}$. The potential shifts by $b^{\rm bc}_{\delta}$ and $b^{\rm bc}_{\theta}$ have not been investigated. 

Using the three-point correlation function,~\citet{Slepian2016:1607.06098v1} constrain the relative velocity parameter $b_{v^2}$ to $b_{v^2} < 0.0097$ ($68\%$ confidence level). When using the $68\%$ confidence levels we find $b_{v^2} = 0.012 \pm 0.015$ when combining the low and high redshift bins. When including the $b_{v^2} < 0$ prior we get tighter constraints of $|b_{v^2}| < 0.007 (< 0.018)$ ($68\%$ and $95\%$ confidence levels). \citet{Slepian2016:1607.06098v1} do not investigate the linear bias parameters $b^{\rm bc}_{\delta}$ and $b^{\rm bc}_{\theta}$.

\citet{1308.1401v2} used the power spectrum monopole to set the constraint $b_{v^2} < 0.033$ ($95\%$ confidence level). However, there is a factor of $3$ difference in the parameterisation, which means that their constraint translates to $b_{v^2} < 0.1$ when using our nomenclature. This constraint is weaker by over one order of magnitude compared to our result. Part of the reason for our much tighter constraint is the increase in survey area between BOSS DR9 (used in~\citealt{1308.1401v2}) and BOSS DR12 (used in this work). Another reason for the improved constraints is the advection term, which is significantly contributing to our parameter constraint and which has not been included in the analysis of~\citet{1308.1401v2}. Note that our results are not depending significantly on the inclusion of the quadrupole.

\citet{1308.1401v2} also pointed out that the relative velocity effect might have an enhanced signature in the cross-correlation of two different galaxy samples. The idea is that one sample contains old galaxies, which formed early and retained the relative velocity effect, while the second sample contains young galaxies which will have a smaller (or no) relative velocity effect. Such an analysis was performed in~\citet{Beutler2015:1506.03900v2} using the BOSS and WiggleZ galaxies. The BOSS sample contains mainly old LRG galaxies, which should carry a stronger relative velocity effect, compared to the ELG galaxies observed in WiggleZ. However, no relative velocity effect was detected and the best obtained constraint was $-0.086 < b_{v^2} < 0.062$ ($68\%$ confidence level). These constraints use the same nomenclature as \citet{1308.1401v2} and hence have to be multiplied by a factor of $3$ before being compared to our constraints. Given that BOSS and WiggleZ overlap only in about $8\%$ of the total BOSS sky coverage, the cosmic volume available for this study was significantly smaller than BOSS alone. This analysis also did not include the advection term.

Finally we note that our measurement of $b_1 \approx 2$ is in good agreement with other studies on the BOSS power spectrum (e.g.~\citealt{Gil-Marin2016:1606.00439v2}), while~\citet{Slepian2016:1607.06098v1} found a smaller value of $b_1 = 1.776\pm0.020$. The tension likely comes from the fact that the model of~\citet{Slepian2016:1607.06098v1} did not include the tidal tensor bias, which can increase $b_1$ to $2.069\pm 0.083$, which is consistent with our measurement.

\section{Conclusion}
\label{sec:conclusion}

We analysed the BOSS DR12 power spectrum multipoles using a power spectrum model for the relative velocity effect. We derive all redshift-space 1-loop terms for the relative velocity, extending models used in previous analysis (see appendix~\ref{app:PTterms}). For the first time we include the advection terms as suggested in~\citet{1510.03554v2}. An analysis without the advection term is presented in~\citet{1308.1401v2}. Besides the relative velocity parameter $b_{v^2}$, we also include the linear density and velocity divergence terms $b^{\rm bc}_{\delta}$ and $b^{\rm bc}_{\theta}$. Our main results can be summarised as follows:
\begin{itemize}
\item We extend the redshift-space clustering model of~\citet{Beutler2014:1312.4611v2, Beutler2016:1607.03150v1} to include all relative velocity terms up to second order in $b_{v^2}$ and linear order in $b^{\rm bc}_{\delta}$ and $b^{\rm bc}_{\theta}$.
\item Using 2 sets of N-body simulations and the BOSS DR12 Multidark Patchy mock catalogues we detect biases in the three relative velocity parameters of up to $2\sigma$ in $b^{\rm bc}_{\theta}$ and $\sim 1\sigma$ in $b_{v^2}$ and $b^{\rm bc}_{\delta}$, indicating shortcomings of our power spectrum model. We correct the measurements by these biases but note that our model for the power spectrum does require further improvement. These biases should be kept in mind when using our constraints.
\item Our data does not support a detection of the relative velocity effect in any of the three relative velocity parameters. Combining the low and high redshift bins, we found limits of $b_{v^2} = 0.012 \pm 0.015 (\pm 0.031)$, $b^{\rm bc}_{\delta} = -1.0 \pm 2.5 (\pm 6.2)$ and $b^{\rm bc}_{\theta} = -114 \pm 55 (\pm 175)$ with $68\%$ ($95\%$) confidence levels. Including a prior of $b_{v^2} < 0$ motivated by treating the relative velocity effect as a pure suppression effect, our constraint on $b_{v^2}$ tightens to $|b_{v^2}| < 0.018$ ($95\%$ confidence levels).
\item Using the BOSS DR12 Fourier-space pipelines for BAO and RSD analysis we quantify the potential systematic uncertainties in the BAO scale and RSD parameter due to the three relative velocity contributions. Our constraints limit the potential systematic shift in $D_A(z)$, $H(z)$ and $f\sigma_8$, due to the relative velocity effect to $1\%$, $0.8\%$ and $2\%$, respectively. Given the current uncertainties on the BAO measurements of BOSS these shifts correspond to $0.53\sigma$, $0.50\sigma$ and $0.22\sigma$ for $D_A(z)$, $H(z)$ and $f\sigma_8$, respectively.
\end{itemize}
In our analysis we did not make use of density field reconstruction, which can significantly improve the BAO signal. Right now we do not have a good model for the broadband shape of the power spectrum post-reconstruction due to the complicated impact of the reconstruction procedure. We therefore leave such investigations for future work. 

\section*{Acknowledgments}

FB would like to thanks Fabian Schmidt for help with the implementation of the $b^{\rm bc}_{\delta}$ and $b^{\rm bc}_{\theta}$ terms as well as valuable comments to this manuscript. FB would also like to thank Jonathan Blazek, Andreu Font-Ribera, Thomas Tram and Shun Saito for fruitful discussions. FB acknowledges support from the UK Space Agency through grant ST/N00180X/1. US is is supported by NASA grant NNX15AL17G. 

\bibliography{BOSS_DR12_relvel_bibtex}{}
\newpage

\appendix
\numberwithin{equation}{section}

\twocolumn[{%

\section{Perturbative terms for the power spectrum model}
\label{app:PTterms}

Our power spectrum model is given by 
\begin{equation}
\begin{split}
P_g(k,\mu) &= \exp\left\{-(fk\mu\sigma_v)^2\right\}\Bigg[P_{\rm g, NL}(k,\mu) + 2b_1b^{\rm bc}_{\delta}P_{\delta\delta_{\rm bc}} + 2b_1b^{\rm bc}_{\theta}P_{\delta\theta{\rm bc}}\\
&+ b_1b_{v^2}\left[P_{\delta|v^2}(k) + P_{\rm adv|\delta}(k)\right] + b_2b_{v^2}P_{\delta^2|v^2}(k) + b_sb_{v^2}P_{s^2|v^2}(k) + b_{v^2}^2P_{v^2|v^2}(k)\\
&- 2f\mu^2\left[b_1b_{v^2}P_{\delta|v^2v_{\parallel}}(k) + b_{v^2}\left(P_{v^2|v_{\parallel}}(k) + P_{\rm adv|v_{\parallel}}(k)\right) + b_1 b_{v^2}P_{v^2|\delta v_{\parallel}}(k)\right]\\
&+ f^2\mu^2b_{v^2}\left[\mu^2P_{v_{\parallel}|v^2v_{\parallel}}(k) - I_{1}(k) - \mu^2I_{2}(k)\right]\Bigg].
\end{split}
\label{eq:psmodel2} 
\end{equation}
The non-linear power spectrum model, $P_{\rm NL}(k,\mu)$ is given by
\begin{equation}
\begin{split}
P_{\rm g, NL}(k,\mu) &= P_{{\rm g},\delta|\delta}(k) + 2f\mu^2P_{{\rm g},\delta|\theta}(k) + f^2\mu^4P_{\theta|\theta}(k) + b_1^3A(k,\mu,\beta) + b_1^4B(k,\mu,\beta), 
\end{split}
\end{equation}
where
\begin{align}
\begin{split}
P_{{\rm g},\delta|\delta}(k) &= b_1^2P_{\delta|\delta}(k) + b_2b_1 P_{\delta|\delta^2}(k) + b_{s}b_1P_{\delta|s^2}(k) + 2b_{\rm 3nl}b_1\sigma_3^2(k)P^{\rm lin}_{\rm m}(k)\\
& + b^2_2P_{\delta^2|\delta^2}(k) + b_2b_{s}P_{\delta^2|s^2}(k) + b^2_{s}P_{s^2|s^2}(k) + N,
\end{split}\\
P_{{\rm g},\delta|\theta}(k) &= b_1P_{\delta|\theta}(k) + b_2P_{\theta|\delta^2}(k) + b_{s}P_{\theta|s^2}(k) + b_{\rm 3nl}\sigma_3^2(k)P^{\rm lin}_{\rm m}(k).
\label{eq:41}
\end{align}
The standard density and velocity terms are given by
\begin{align}
P_{\delta|\delta^2}(k) &= 2\int \frac{d^3q}{(2\pi)^3}P^{\rm lin}_{\rm m}(q)P^{\rm lin}_{\rm m}(k-q) F_{2}(\vc{q},\vc{k-q}),\\
P_{\theta|\delta^2}(k) &= \int \frac{d^3q}{(2\pi)^3}P^{\rm lin}_{\rm m}(q)P^{\rm lin}_{\rm m}(k-q) G_{2}(\vc{q},\vc{k-q}),\\
P_{\delta|s^2}(k) &= 2\int \frac{d^3q}{(2\pi)^3}P^{\rm lin}_{\rm m}(q)P^{\rm lin}_{\rm m}(k-q) F_{2}(\vc{q},\vc{k-q})S_{2}(\vc{q},\vc{k-q}),\\
P_{\theta|s^2}(k) &= \int \frac{d^3q}{(2\pi)^3}P^{\rm lin}_{\rm m}(q)P^{\rm lin}_{\rm m}(k-q) G_{2}(\vc{q},\vc{k-q})S_{2}(\vc{q},\vc{k-q}),\\
P_{\delta^2|\delta^2}(k) &= \frac{1}{2}\int \frac{d^3q}{(2\pi)^3}P^{\rm lin}_{\rm m}(q)\Big[P^{\rm lin}_{\rm m}(k-q) - P^{\rm lin}_{\rm m}(q)\Big],\\
P_{\delta^2|s^2}(k) &= -\int \frac{d^3q}{(2\pi)^3}P^{\rm lin}_{\rm m}(q)\Big[\frac{2}{3}P^{\rm lin}_{\rm m}(q) - P^{\rm lin}_{\rm m}(k-q)S_{2}(\vc{q},\vc{k-q})\Big],\\
P_{s^2|s^2}(k) &= -\frac{1}{2}\int \frac{d^3q}{(2\pi)^3}P^{\rm lin}_{\rm m}(q)\Big[\frac{4}{9}P^{\rm lin}_{\rm m}(q) - P^{\rm lin}_{\rm m}(k-q)S_{2}(\vc{q},\vc{k-q})^2\Big],\\
\sigma_{3}^{2}(k) &= \frac{105}{16}\int \frac{d^3 q}{(2\pi)^{3}}P^{\rm lin}_{\rm m}(q)
                \left[  D_{2}(-\vc{q},\vc{k})S_{2}(\vc{q}, \vc{k-q}) + \frac{8}{63}  \right].
\label{eq:nlterms}
\end{align}
The additional relative velocity terms without redshift-space distortions are 
\begin{align}
P_{\rm adv|\delta}(k) &= \frac{4}{3}T_{v}(k)kP^{\rm lin}_{\rm m}(k)L_s,\\
P_{\delta|v^2}(k) &= 4\int \frac{d^3\vc{q}}{(2\pi)^3}P^{\rm lin}_{\rm m}(q)P^{\rm lin}_{\rm m}(k-q)F_2(\vc{q},\vc{k-q})G_u(\vc{q},\vc{k-q}) \mu(\vc{q},\vc{k-q}),\\
P_{\delta^2|v^2}(k) &= 2\int \frac{d^3\vc{q}}{(2\pi)^3}P^{\rm lin}_{\rm m}(q)\left[P^{\rm lin}_{\rm m}(k-q)\mu(\vc{q},\vc{k-q})G_u(\vc{q},\vc{k-q}) + P^{\rm lin}_{\rm m}(q)G_u(\vc{q},\vc{q})\right],\\
P_{s^2|v^2}(k) &= 2\int \frac{d^3\vc{q}}{(2\pi)^3}P^{\rm lin}_{\rm m}(q)\left[P^{\rm lin}_{\rm m}(k-q)S_2(\vc{q},\vc{k-q})\mu(\vc{q},\vc{k-q})G_u(\vc{q},\vc{k-q}) + \frac{2}{3}P^{\rm lin}_{\rm m}(q)G_u(\vc{q},\vc{q}) \right],\\
P_{v^2|v^2}(k) &= 2\int \frac{d^3\vc{q}}{(2\pi)^3}P^{\rm lin}_{\rm m}(q)\left[ P^{\rm lin}_{\rm m}(k-q)\mu^2(\vc{q},\vc{k-q})G^2_u(\vc{q},\vc{k-q}) - P^{\rm lin}_{\rm m}(q)G^2_u(\vc{q},\vc{q})\right]
\end{align}
with $\mu(\vc{k_1},\vc{k_2}) = \frac{\vc{k_1}\cdot\vc{k_2}}{k_1k_2}$ and
\begin{equation}
L_s = \int \frac{k\,dk}{2\pi^2}T_{v}(k)P_{\rm lin}(k).
\end{equation}
}]

\twocolumn[{%

The relative velocity redshift-space distortion terms are 
\begin{align}
P_{\delta|v^2v_{\parallel}}(k) &= \frac{2}{3}T_{v}(k)kP_{\rm lin}(k)L_s = \frac{1}{2}P_{\rm adv|\delta}(k),\\
P_{v^2|v_{\parallel}}(k) &= 2\int \frac{d^3\vc{q}}{(2\pi)^3}\frac{k\mu - q}{\sqrt{k^2 - 2kq\mu + q^2}} P_{\rm lin}(q)P_{\rm lin}(k-q)G_2(\vc{q},\vc{k-q})G_u(\vc{q},\vc{k-q}),\\
P_{\text{adv} |v_{\parallel}}(k) &= -\frac{2}{3}T_{v}(k)kP_{\rm lin}(k)L_s = -\frac{1}{2}P_{\rm adv|\delta}(k) = -P_{\delta|v^2v_{\parallel}}(k),\\
P_{v^2|\delta v_{\parallel}}(k) &= 2\int \frac{d^3\vc{q}}{(2\pi)^3}\frac{k\mu(k\mu - q)}{q\sqrt{k^2 - 2kq\mu + q^2}} P_{\rm lin}(q)P_{\rm lin}(k-q)G_u(\vc{q},\vc{k-q}),\\
P_{v_{\parallel}|v^2v_{\parallel}}(k) &= -\frac{4}{3}T_{v}(k)kP_{\rm lin}(k)L_s = -P_{\rm adv|\delta}(k),\\
P_{v^2| v^2_{\parallel}}(k) &=   I_{1}(k) + \mu^2I_{2}(k)
\end{align}
with
\begin{align}
I_{1}(k) &= k^2\int \frac{d^3\vc{q}}{(2\pi)^3}\frac{k^2(1-\mu^2)(q-k\mu)}{\left[k^2 - 2kq\mu + q^2\right]^{3/2}} G_u(\vc{q},\vc{k-q})P_{\rm lin}(q)P_{\rm lin}(k-q),\\
I_{2}(k) &= k^2\int \frac{d^3\vc{q}}{(2\pi)^3}\frac{k^2(2k^2\mu^2 - k(3\mu^3 + \mu)q + (3\mu^2 - 1)q^2)}{q\left[k^2 - 2kq\mu + q^2\right]^{3/2}} G_u(\vc{q},\vc{k-q})P_{\rm lin}(q)P_{\rm lin}(k-q).
\end{align}
The symmetrised $2$nd-order PT kernels, $F_{2}$, $G_{2}$, $S_{2}$ and $G_u$ are given by
\begin{align}
F_2(\vc{k_1},\vc{k_2}) &= \frac{5}{7} + \frac{2}{7}\left(\frac{\vc{k_1}\cdot \vc{k_2}}{k_1k_2}\right)^2 + \frac{\vc{k_1}\cdot \vc{k_2}}{2}\left(\frac{1}{k_1^2} + \frac{1}{k_2^2}\right),\\
G_{2}(\vc{k_1},\vc{k_2}) &= \frac{3}{7} + \frac{\vc{k_1}\cdot \vc{k_2}}{2}
\left(\frac{1}{k^2_1} + \frac{1}{k^2_2}\right) + \frac{4}{7}\left(\frac{\vc{k_1}\cdot \vc{k_2}}{k_1k_2}\right)^2,\\
S_2(\vc{k_1},\vc{k_2}) &= \left(\frac{\vc{k_1}\cdot\vc{k_2}}{k_1k_2}\right)^2 - \frac{1}{3},\\
D_{2}(\vc{k_1},\vc{k_2}) &= \frac{2}{7}\left[S_{2}(\vc{k_1},\vc{k_2}) - \frac{2}{3}\right],\\
G_u(\vc{k_1},\vc{k_2}) &= -T_v(k_1)T_{v}(k_2).
\end{align}
}]

\section{Tables with fitting results}

\begin{landscape}

\begin{table}
   \begin{center}
      \caption{Fitting results for the mean of the $20$ runA simulations in redshift-space. We show the results for the three individual relative velocity parameters and for varying all three parameters simultaneously. The covariance matrix is derived from the Multidark Patchy mock catalogues scaled according to the volume. We show $68\%$ confidence levels for most parameters, but also include the $95\%$ confidence levels in parentheses for the relative velocity parameters.}
      \begin{tabular}{lllllllllll}
         \hline
         \multicolumn{11}{c}{Test on mean of the 20 runA redshift-space mocks}\\
         & \multicolumn{2}{c}{no rel. vel.} & \multicolumn{2}{c}{only $b_{v^2}$} & \multicolumn{2}{c}{only $b_{\delta}$} & \multicolumn{2}{c}{only $b_{\theta}$} & \multicolumn{2}{c}{$b_{v^2} + b_{\delta}$}\\
         \hline
          & max. like. & mean & max. like. & mean & max. like. & mean & max. like. & mean & max. like. & mean\\
          $\alpha_{\perp}$  & $ 1 $ & $ 1 $ & $ 1 $ & $ 1 $ & $ 1 $ & $ 1 $ & $ 1 $ & $ 1 $ & $ 1 $ & $ 1 $ \\
          $\alpha_{\parallel}$  & $ 1 $ & $ 1 $ & $ 1 $ & $ 1 $ & $ 1 $ & $ 1 $ & $ 1 $ & $ 1 $ & $ 1 $ & $ 1 $ \\
          $f\sigma_8$  & $ 0.455 $ & $ 0.455 $ & $ 0.455 $ & $ 0.455 $ & $ 0.455 $ & $ 0.455 $ & $ 0.455 $ & $ 0.455 $ & $ 0.455 $ & $ 0.455 $ \\
          $b_v[10^{-3}]$  & $ 0 $ & $ 0 $ & $ 21.9 $ & $ 22.2 \pm 6.8 (\pm 14) $ & $ 0 $ & $ 0 $ & $ 0 $ & $ 0 $ & $ 7 $ & $ 16 \pm 21 (\pm 44) $ \\
          $b_{\delta}$  & $ 0 $ & $ 0 $ & $ 0 $ & $ 0 $ & $ -3.6 $ & $ -3.5 \pm 1.1 (\pm 2.1) $ & $ 0 $ & $ 0 $ & $ -2.7 $ & $ -1.4 \pm 3.4 (\pm 6.9) $ \\
          $b_{\theta}$  & $ 0 $ & $ 0 $ & $ 0 $ & $ 0 $ & $ 0 $ & $ 0 $ & $ 142 $ & $ 147 \pm 51 (^{+ 170 }_{- 98 }) $ & $ 0 $ & $ 0 $ \\
          \hline
          $b_1\sigma_8$  & $ 1.219 $ & $ 1.220 ^{+ 0.013 }_{- 0.018 } $ & $ 1.2198 $ & $ 1.2230 ^{+ 0.0069 }_{- 0.0096 } $ & $ 1.2213 $ & $ 1.2208 \pm 0.0067 $ & $ 1.2173 $ & $ 1.2178 ^{+ 0.0081 }_{- 0.011 } $ & $ 1.2218 $ & $ 1.2200 ^{+ 0.0083 }_{- 0.011 } $ \\
          $b_2\sigma_8$  & $ 0.12 $ & $ 0.32 ^{+ 0.70 }_{- 0.47 } $ & $ 0.37 $ & $ 0.61 \pm 0.49 $ & $ 0.68 $ & $ 0.71 \pm 0.37 $ & $ 0.38 $ & $ 0.48 ^{+ 0.46 }_{- 0.36 } $ & $ 0.66 $ & $ 0.60 \pm 0.60 $ \\
          N  & $ 274.0 $ & $ -100 \pm 1200 $ & $ -80 $ & $ -410 ^{+ 760 }_{- 580 } $ & $ -1090 $ & $ -1150 \pm 730 $ & $ -350 $ & $ -550 \pm 860 $ & $ -940 $ & $ -550 ^{+ 1500 }_{- 1000 } $ \\
          $\sigma_v$  & $ -4.80 $ & $ -4.85 \pm 0.18 $ & $ -4.83 $ & $ -4.90 ^{+ 0.15 }_{- 0.11 } $ & $ -4.82 $ & $ -4.83 \pm 0.11 $ & $ -4.85 $ & $ -4.87 \pm 0.14 $ & $ -4.84 $ & $ -4.85 \pm 0.16 $ \\
         \hline
      \end{tabular}
      \label{tab:runA}
   \end{center}
\end{table}

\begin{table}
   \begin{center}
      \caption{Same as table~\ref{tab:runA} but for the $10$ runPB simulations. We show $68\%$ confidence levels for most parameters, but also include the $95\%$ confidence levels in parenthesis for the relative velocity parameters.}
      \begin{tabular}{lllllllllll}
         \hline
         \multicolumn{11}{c}{Test on mean of the 10 runPB redshift-space mocks}\\
         & \multicolumn{2}{c}{no rel. vel.} & \multicolumn{2}{c}{only $b_{v^2}$} & \multicolumn{2}{c}{only $b_{\delta}$} & \multicolumn{2}{c}{only $b_{\theta}$} & \multicolumn{2}{c}{$b_{v^2} + b_{\delta}$}\\
         \hline
          & max. like. & mean & max. like. & mean & max. like. & mean & max. like. & mean & max. like. & mean\\
          $\alpha_{\perp}$  & $ 1 $ & $ 1 $ & $ 1 $ & $ 1 $ & $ 1 $ & $ 1 $ & $ 1 $ & $ 1 $ & $ 1 $ & $ 1 $ \\
          $\alpha_{\parallel}$  & $ 1 $ & $ 1 $ & $ 1 $ & $ 1 $ & $ 1 $ & $ 1 $ & $ 1 $ & $ 1 $ & $ 1 $ & $ 1 $ \\
          $f\sigma_8$  & $ 0.472 $ & $ 0.472 $ & $ 0.472 $ & $ 0.472 $ & $ 0.472 $ & $ 0.472 $ & $ 0.472 $ & $ 0.472 $ & $ 0.472 $ & $ 0.472 $ \\
          $b_v[10^{-3}]$  & $ 0 $ & $ 0 $ & $ 19 $ & $ 20 \pm 11 (\pm 21) $ & $ 0 $ & $ 0 $ & $ 0 $ & $ 0 $ & $ 48.8 $ & $ 45 \pm 26 (\pm 50) $ \\
          $b_{\delta}$  & $ 0 $ & $ 0 $ & $ 0 $ & $ 0 $ & $ -2.2 $ & $ -2.3 \pm 1.5( \pm 3.0) $ & $ 0 $ & $ 0 $ & $ 5.0 $ & $ 4.2 \pm 4.1 (\pm 8.1) $ \\
          $b_{\theta}$  & $ 0 $ & $ 0 $ & $ 0 $ & $ 0 $ & $ 0 $ & $ 0 $ & $ 82 $ & $ 77 \pm 63 (\pm 120) $ & $ 0 $ & $ 0 $ \\
          \hline
          $b_1\sigma_8$  & $ 1.268 $ & $ 1.268 ^{+ 0.013 }_{- 0.016 } $ & $ 1.268 $ & $ 1.2684 \pm 0.0092 $ & $ 1.2683 $ & $ 1.2684 ^{+ 0.0092 }_{- 0.012 } $ & $ 1.267 $ & $ 1.268 ^{+ 0.011 }_{- 0.014 } $ & $ 1.262 $ & $ 1.266 \pm 0.012 $ \\
          $b_2\sigma_8$  & $ 0.42 $ & $ 0.54 ^{+ 0.59 }_{- 0.44 } $ & $ 0.65 $ & $ 0.79 \pm 0.48 $ & $ 0.72 $ & $ 0.79 \pm 0.46 $ & $ 0.56 $ & $ 0.70 ^{+ 0.57 }_{- 0.45 } $ & $ 0.21 $ & $ 0.50 ^{+ 0.76 }_{- 0.54 } $ \\
          N  & $ -500 $ & $ -760 \pm 1100 $ & $ -690 $ & $ -760 \pm 730 $ & $ -1200 $ & $ -1360 ^{+ 1000 }_{- 810 } $ & $ -820 $ & $ -1120 ^{+ 1100 }_{- 880 } $ & $ 950 $ & $ 520 \pm 1400 $ \\
          $\sigma_v$  & $ -5.55 $ & $ -5.57 \pm 0.15 $ & $ -5.57 $ & $ -5.60 \pm 0.12 $ & $ -5.54 $ & $ -5.55 \pm 0.13 $ & $ -5.57 $ & $ -5.59 \pm 0.14 $ & $ -5.59 $ & $ -5.62 ^{+ 0.13 }_{- 0.17 } $ \\
         \hline
      \end{tabular}
      \label{tab:runPB}
   \end{center}
\end{table}

\end{landscape}

\begin{landscape}

\begin{table}
   \begin{center}
      \caption{Fitting results for our relative velocity model using the low redshift bin ($0.2 < z < 0.5$) of the Multidark Patchy mock catalogues including the NGC window function. We show $68\%$ confidence levels for most parameters, but also include the $95\%$ confidence levels in parenthesis for the relative velocity parameters.}
      \begin{tabular}{lllllllllll}
         \hline
         \multicolumn{11}{c}{Test on mean of the 2045 Multidark Patchy mocks for the redshift range $0.2 < z < 0.5$.}\\
         & \multicolumn{2}{c}{no rel. vel.} & \multicolumn{2}{c}{only $b_{v^2}$} & \multicolumn{2}{c}{only $b_{\delta}$} & \multicolumn{2}{c}{only $b_{\theta}$} & \multicolumn{2}{c}{$b_{v^2} + b_{\delta}$}\\
         \hline
          & max. like. & mean & max. like. & mean & max. like. & mean & max. like. & mean & max. like. & mean\\
          $\alpha_{\perp}$  & $ 1 $ & $ 1 $ & $ 1 $ & $ 1 $ & $ 1 $ & $ 1 $ & $ 1 $ & $ 1 $ & $ 1 $ & $ 1 $ \\
          $\alpha_{\parallel}$  & $ 1 $ & $ 1 $ & $ 1 $ & $ 1 $ & $ 1 $ & $ 1 $ & $ 1 $ & $ 1 $ & $ 1 $ & $ 1 $ \\
          $f\sigma_8$  & $ 0.484 $ & $ 0.484 $ & $ 0.484 $ & $ 0.484 $ & $ 0.484 $ & $ 0.484 $ & $ 0.484 $ & $ 0.484 $ & $ 0.484 $ & $ 0.484 $ \\
          $b_v[10^{-3}]$  & $ 0 $ & $ 0 $ & $ 29.1 $ & $ 29.8 \pm 5.0 (\pm 9.6) $ & $ 0 $ & $ 0 $ & $ 0 $ & $ 0 $ & $ 36 $ & $ 34 \pm 12 (\pm 22) $ \\
          $b_{\delta}$  & $ 0 $ & $ 0 $ & $ 0 $ & $ 0 $ & $ -4.96 $ & $ -4.78 \pm 0.78 (\pm 1.6) $ & $ 0 $ & $ 0 $ & $ 1.2 $ & $ 0.5 \pm 2.0 (\pm 4.2) $ \\
          $b_{\theta}$  & $ 0 $ & $ 0 $ & $ 0 $ & $ 0 $ & $ 0 $ & $ 0 $ & $ 187.2 $ & $ 187.0 \pm 6.8 (\pm 9.6) $ & $ 0 $ & $ 0 $ \\
\hline
          $b_{1}^{\rm NGC}\sigma_8$  & $ 1.347 $ & $ 1.345 \pm 0.010 $ & $ 1.3489 $ & $ 1.345 \pm 0.010 $ & $ 1.3547 $ & $ 1.3521 ^{+ 0.0076 }_{- 0.0095 } $ & $ 1.348 $ & $ 1.344 \pm 0.033 $ & $ 1.3392 $ & $ 1.3432 \pm 0.0093 $ \\
          $b_{1}^{\rm SGC}\sigma_8$  & $ 1.344 $ & $ 1.348 \pm 0.016 $ & $ 1.3591 $ & $ 1.359 ^{+ 0.011 }_{- 0.014 } $ & $ 1.3585 $ & $ 1.3600 ^{+ 0.0092 }_{- 0.013 } $ & $ 1.343 $ & $ 1.348 \pm 0.031 $ & $ 1.352 $ & $ 1.355 \pm 0.012 $ \\
          $b^{\rm NGC}_2\sigma_8$  & $ 0.17 $ & $ 0.15 \pm 0.20 $ & $ 0.50 $ & $ 0.58 ^{+ 0.30 }_{- 0.23 } $ & $ 0.90 $ & $ 0.82 \pm 0.26 $ & $ 0.46 $ & $ 0.5 \pm 1.5 $ & $ 0.27 $ & $ 0.44 ^{+ 0.45 }_{- 0.25 } $ \\
          $b^{\rm SGC}_2\sigma_8$  & $ -0.01 $ & $ 0.08 \pm 0.28 $ & $ 0.60 $ & $ 0.67 \pm 0.38 $ & $ 0.80 $ & $ 0.89 \pm 0.34 $ & $ 0.2 $ & $ 0.2 \pm 1.3 $ & $ 0.41 $ & $ 0.57 ^{+ 0.45 }_{- 0.33 } $ \\
          N$^{\rm NGC}$  & $ -470 $ & $ -420 \pm 680 $ & $ -980 $ & $ -1040 \pm 530 $ & $ -2550 $ & $ -2360 \pm 600 $ & $ -1340 $ & $ -1330 \pm 320 $ & $ -260 $ & $ -600 \pm 840 $ \\
          N$^{\rm SGC}$  & $ 210 $ & $ -60 \pm 980 $ & $ -1150 $ & $ -1190 \pm 670 $ & $ -2260 $ & $ -2460 ^{+ 870 }_{- 660 } $ & $ -510 $ & $ -510 \pm 330 $ & $ -530 $ & $ -930 \pm 870 $ \\
          $\sigma^{\rm NGC}_v$  & $ 5.89 $ & $ 5.87 \pm 0.11 $ & $ 5.94 $ & $ 5.96 \pm 0.11 $ & $ 5.94 $ & $ 5.91 \pm 0.10 $ & $ 5.95 $ & $ 5.89 \pm 0.66 $ & $ 5.91 $ & $ 5.94 ^{+ 0.12 }_{- 0.09 } $ \\
          $\sigma^{\rm SGC}_v$  & $ 5.88 $ & $ 5.91 \pm 0.14 $ & $ 6.06 $ & $ 6.08 \pm 0.14 $ & $ 5.99 $ & $ 6.01 \pm 0.13 $ & $ 5.92 $ & $ 5.97 \pm 0.68 $ & $ 6.03 $ & $ 6.06 \pm 0.14 $ \\
         \hline
      \end{tabular}
      \label{tab:patchy_z1_red}
   \end{center}
\end{table}

\begin{table}
   \begin{center}
      \caption{Fitting results for our relative velocity model using the high redshift bin of the Multidark Patchy mock catalogues including the NGC window function. We show $68\%$ confidence levels for most parameters, but also include the $95\%$ confidence levels in parenthesis for the relative velocity parameters.}
      \begin{tabular}{lllllllllll}
         \hline
         \multicolumn{11}{c}{Test on mean of the 2045 Multidark Patchy redshift-space mocks in the redshift range $0.5 < z < 0.75$}\\
         & \multicolumn{2}{c}{no rel. vel.} & \multicolumn{2}{c}{only $b_{v^2}$} & \multicolumn{2}{c}{only $b_{\delta}$} & \multicolumn{2}{c}{only $b_{\theta}$} & \multicolumn{2}{c}{$b_{v^2} + b_{\delta}$}\\
         \hline
          & max. like. & mean & max. like. & mean & max. like. & mean & max. like. & mean & max. like. & mean\\
          $\alpha_{\perp}$  & $ 1 $ & $ 1 $ & $ 1 $ & $ 1 $ & $ 1 $ & $ 1 $ & $ 1 $ & $ 1 $ & $ 1 $ & $ 1 $ \\
          $\alpha_{\parallel}$  & $ 1 $ & $ 1 $ & $ 1 $ & $ 1 $ & $ 1 $ & $ 1 $ & $ 1 $ & $ 1 $ & $ 1 $ & $ 1 $ \\
          $f\sigma_8$  & $ 0.478 $ & $ 0.478 $ & $ 0.478 $ & $ 0.478 $ & $ 0.478 $ & $ 0.478 $ & $ 0.478 $ & $ 0.478 $ & $ 0.478 $ & $ 0.478 $ \\
          $b_v[10^{-3}]$  & $ 0 $ & $ 0 $ & $ 27.6 $ & $ 27.0 ^{+ 6.2}_{- 7.9 }(^{+ 19}_{- 22})  $ & $ 0 $ & $ 0 $ & $ 0 $ & $ 0 $ & $ 40.3 $ & $ 40.3 \pm 9.4 (\pm 20) $ \\
          $b_{\delta}$  & $ 0 $ & $ 0 $ & $ 0 $ & $ 0 $ & $ -3.44 $ & $ -3.47 \pm 0.66 (\pm 1.3) $ & $ 0 $ & $ 0 $ & $ 2.1 $ & $ 2.0 \pm 1.4 (^{+ 3.4 }_{- 2.6 }) $ \\
          $b_{\theta}$  & $ 0 $ & $ 0 $ & $ 0 $ & $ 0 $ & $ 0 $ & $ 0 $ & $ 191.9 $ & $ 192.5 \pm 6.5 (\pm 9.4) $ & $ 0 $ & $ 0 $ \\
\hline
          $b_{1}^{\rm NGC}\sigma_8$  & $ 1.3309 $ & $ 1.3320 \pm 0.0087 $ & $ 1.3266 $ & $ 1.3272 \pm 0.0064 $ & $ 1.3244 $ & $ 1.3250 ^{+ 0.0058 }_{- 0.0075 } $ & $ 1.324 $ & $ 1.326 \pm 0.032 $ & $ 1.3255 $ & $ 1.3266 \pm 0.0067 $ \\
          $b_{1}^{\rm SGC}\sigma_8$  & $ 1.308 $ & $ 1.311 \pm 0.012 $ & $ 1.3175 $ & $ 1.3169 \pm 0.0066 $ & $ 1.3116 $ & $ 1.3138 ^{+ 0.0078 }_{- 0.010 } $ & $ 1.312 $ & $ 1.310 \pm 0.036 $ & $ 1.3143 $ & $ 1.3161 ^{+ 0.0080 }_{- 0.011 } $ \\
          $b^{\rm NGC}_2\sigma_8$  & $ 0.55 $ & $ 0.60 \pm 0.29 $ & $ 0.83 $ & $ 0.85 \pm 0.32 $ & $ 0.82 $ & $ 0.92 \pm 0.25 $ & $ 0.8 $ & $ 1.0 \pm 1.4 $ & $ 0.67 $ & $ 0.74 \pm 0.33 $ \\
          $b^{\rm SGC}_2\sigma_8$  & $ 0.19 $ & $ 0.27 ^{+ 0.39 }_{- 0.29 } $ & $ 0.86 $ & $ 0.84 ^{+ 0.39 }_{- 0.28 } $ & $ 0.70 $ & $ 0.85 \pm 0.32 $ & $ 0.8 $ & $ 1.0 \pm 1.3 $ & $ 0.59 $ & $ 0.68 \pm 0.41 $ \\
          N$^{\rm NGC}$  & $ -999 $ & $ -1110 \pm 620 $ & $ -943 $ & $ -900 \pm 350 $ & $ -1570 $ & $ -1710 \pm 420 $ & $ -1530 $ & $ -1520 \pm 330 $ & $ -290 $ & $ -360 \pm 600 $ \\
          N$^{\rm SGC}$  & $ 0 $ & $ -200 \pm 800 $ & $ -873 $ & $ -800 \pm 320 $ & $ -1220 $ & $ -1430 \pm 610 $ & $ -1270 $ & $ -1240 \pm 310 $ & $ -100 $ & $ -170 \pm 690 $ \\
          $\sigma^{\rm NGC}_v$  & $ 5.750 $ & $ 5.760 \pm 0.097 $ & $ 5.758 $ & $ 5.762 \pm 0.085 $ & $ 5.678 $ & $ 5.697 \pm 0.084 $ & $ 5.76 $ & $ 5.81 \pm 0.69 $ & $ 5.777 $ & $ 5.788 \pm 0.082 $ \\
          $\sigma^{\rm SGC}_v$  & $ 5.69 $ & $ 5.71 \pm 0.13 $ & $ 5.83 $ & $ 5.86 \pm 0.10 $ & $ 5.71 $ & $ 5.73 \pm 0.12 $ & $ 5.81 $ & $ 5.83 \pm 0.65 $ & $ 5.82 $ & $ 5.842 \pm 0.098 $ \\
         \hline
      \end{tabular}
      \label{tab:patchy_z3_red}
   \end{center}
\end{table}

\end{landscape}

\label{lastpage}

\end{document}